\documentclass{aa}				
\usepackage{graphicx}
\usepackage{txfonts}

\usepackage{amsmath}
\usepackage{url}

\usepackage{natbib,twoopt}

\bibpunct{(}{)}{;}{a}{}{,} 

\usepackage[breaklinks=true]{hyperref} 
\bibpunct{(}{)}{;}{a}{}{,} 
\makeatletter
\newcommandtwoopt{\citeads}[3][][]{\href{http://adsabs.harvard.edu/abs/#3}%
{\def\hyper@linkstart##1##2{}%
\let\hyper@linkend\@empty\citealp[#1][#2]{#3}}}
\newcommandtwoopt{\citepads}[3][][]{\href{http://adsabs.harvard.edu/abs/#3}%
{\def\hyper@linkstart##1##2{}%
\let\hyper@linkend\@empty\citep[#1][#2]{#3}}}
\newcommandtwoopt{\citetads}[3][][]{\href{http://adsabs.harvard.edu/abs/#3}%
{\def\hyper@linkstart##1##2{}%
\let\hyper@linkend\@empty\citet[#1][#2]{#3}}}
\newcommandtwoopt{\citeyearads}[3][][]%
{\href{http://adsabs.harvard.edu/abs/#3}
{\def\hyper@linkstart##1##2{}%
\let\hyper@linkend\@empty\citeyear[#1][#2]{#3}}}
\makeatother

\begin{document}

\title{Correcting systematic polarization effects in Keck LRISp spectropolarimetry to $<$0.05\% }

\titlerunning{Correcting Keck LRISp spectropolarimetry}

\author{David M. Harrington \inst{1,} \inst{2,} \inst{3} \and Svetlana V. Berdyugina \inst{1} \inst{2} \and Oleksii Kuzmychov \inst{1} \and Jeffrey R. Kuhn \inst{4} }
\authorrunning{Harrington et. al.}

\institute{Kiepenheuer-Institut f\"{u}r Sonnenphysik, Sch\"{o}neckstr. 6, D-79104 Freiburg, Germany \\
\and Institute for Astronomy, University of Hawaii, 2680 Woodlawn Drive, Honolulu, HI, 96822, USA  \\
\and Applied Research Labs, University of Hawaii, 2800 Woodlawn Drive, Honolulu, HI, 96822, USA  \\
\and  Institute for Astronomy Maui, University of Hawaii, 34 Ohia Ku St., Pukalani, HI, 96768, USA 
 }

\date{Submitted March, 2015}

\abstract{Spectropolarimetric measurements at moderate spectral resolutions are effective tracers of stellar magnetic fields and circumstellar environments when signal to noise ratios (SNRs) above 2000 can be achieved. The LRISp spectropolarimeter is capable of achieving these SNRs on faint targets with the 10m aperture of the Keck telescope, provided several instrumental artifacts can be suppressed. We describe here several methods to overcome instrumental error sources that are required to achieve these high SNRs on LRISp. We explore high SNR techniques such as defocusing and slit-stepping during integration with high spectral and spatial oversampling.  We find that the instrument flexure and interference fringes introduced by the achromatic retarders create artificial signals at 0.5\% levels in the red channel which mimic real stellar signals and limit the sensitivity and calibration stability of LRISp.  Careful spectral extraction and data filtering algorithms can remove these error sources. For faint targets and long exposures, cosmic ray hits are frequent and present a major limitation to the upgraded deep depletion red-channel CCD. These must be corrected to the same high SNR levels, requiring careful spectral extraction using iterative filtering algorithms. We demonstrate here characterization of these sources of instrumental polarization artifacts and present several methods used to successfully overcome these limitations. We have measured the linear to circular cross-talk and find it to be roughly 5\%, consistent with the known instrument limitations. We show spectropolarimetric signals on brown dwarfs are clearly detectable at 0.2\% amplitudes with sensitivities better than 0.05\% at full spectral sampling in atomic and molecular bands. Future LRISp users can perform high sensitivity observations with high quality calibration when following the described algorithms. }


\maketitle

\section{Introduction}

The Keck telescope on Mauna Kea and the Low-Resolution Imaging Spectrograph LRIS delivers a very large collecting area (10m) to a moderate resolution Cassegrain slit spectrograph \citep{Goodrich:2003kv,1995PASP..107..375O}.  The polarimetric mode, called LRISp, provides a dual-beam mode with the ability to measure circular and linear polarized spectra using achromatic retarders \citep{Goodrich:1995fg,1991PASP..103.1314G} to modulate the incoming polarized light.   

In the field of stellar magnetism, several recent theoretical and observational studies show that resolutions of only a few thousand are required to detect signatures of global fields and star spots.  Magnetic fields in TiO bands have been detected in M-dwarfs. \citep{2008ASPC..384..175B}. Iron hydride (FeH) and chromium hydride (CrH) bands are also observable and modeled to be detectable with high polarimetric sensitivity \citep{Afram:2008kt, Afram:2007iq, Kuzmychov:2013hb}. The 3-dimensional structure of star spots can be constrained with observations in multiple molecular bands \citep{Berdyugina:2011wc}. With LRISp, several scientific investigations are possible provided that signal-to-noise ratios (SNRs) of over 1000 can be delivered as we have been pursuing \citep{Kuzmychov:2014vv, 2013MmSAI..84.1127K}. 

Many galactic sources are highly polarized showing detectable continuum and line polarization effects of up to 20\% \citep{1995AJ....110.2597T,1995ApJ...440..578T,1995ApJ...440..597T, 1995ApJ...440..565T}. LRISp has been used to search these targets achieving roughly 1\% polarimetric sensitivites \citep{2011ApJ...726L..21T}.

To achieve these high SNRs, not only are many photons required, but a thorough calibration and correction of many instrumental artifacts must be performed. Most instruments suffer from several problems that create artifacts in polarimetric data. For example, instrument flexure causes wavelength drifts of a fraction of a pixel between exposures or within the two polarized beams of a single exposure. These small instrumental wavelength shifts mimic signatures from stellar magnetic fields which are also wavelength shifts of spectral lines through the Zeeman effect (c.f. \citet{2013A&A...559A.103B}). A typical polarization calculation requires combining spectra from several exposures. Rotating retarders, as in LRISp and other instruments, can introduce several polarimetric artifacts \citep{2008PASP..120...89H}.  Instrumental wavelength instabilities can become a serious limitation even with shifts as small as 0.1 pixel. The flexure of LRIS was measured in 2011 by telescope staff to be over a pixel \citep{Keck:2011vc} similar to previous LRIS flexure measurements \citep{Keck:2014uc}. 

Polarimetric instruments typically use retarders to modulate the incoming stellar light and a polarizer to function as an {\it analyzer}.  Dual beam instruments utilize polarizing beamsplitters so that orthogonally analyzed polarization states are recorded on the CCD with high instrument throughput and the ability to remove systematic errors. By differencing and / or ratioing spectra recorded with different retarder orientations through the two orthogonally polarized dual beams, several instrument systematic errors can be removed. This is typically called {\it beam swapping}. Several examples are included in these references: \citep{2003isp..book.....D, Bagnulo:2009bz, Tinbergen:2005up, 2013pss2.book..175S, 1993A&A...278..231S, 1974psns.coll.....G}. Dual beam instrumentation with {\it beam swapping} modulation techniques can cancels out several effects to first order. Depending on how the polarization spectra are computed, there are still second-order artifacts remaining from instabilities in time (seeing, pointing jitter, sky transparency) and between the beams (transmissions, flat-fielding, differential aberrations, CCD imperfections, etc).  Since polarization spectra are computed as differences between measured intensities, instrumental artifacts must be carefully removed or mitigated \citep{1996SoPh..164..243K}.

Several kinds of retarders often have small but detectable interference effects, called fringes in some texts and are more generally polarized spectral fringes \citep{2005A&A...434..377C, 2003A&A...401....1S}. Many retarders are manufactured as multiple layers of birefringent materials that produce Fabry-Perot type etaloning which introduces fringes. These fringes are different for both the fast and slow axis orientations, creating spurious instrumental polarization. Some of the {\it super-achromatic} type retarders with multiple layers can have fringes producing spectral intensity modulation of well over 1\% amplitudes.  These fringes are often modeled or removed to some residual error level with various function fits or Fourier filtering techniques. \citep{Aitken:2001ih, Harries:1996vf, 1995ApJ...448L..49A, 2005A&A...434..377C, 2003A&A...401....1S}.

Telescope and instrumental polarization is important because knowledge of the continuum polarization can be used in addition to line polarization as a constraint on circumstellar environments.  Because this dual-beam spectropolarimeter is mounted at Cassegrain focus in a mostly symmetric optical beam, it has quite minimal instrumental polarization. The instrument is mounted roughly 10 arc minutes off-axis showing low polarization induced by the telescope. Most telescopes even in axially symmetric beams show instrumental induced polarization at the 0.1\% level from asymmetries in the optical coatings, oxidation, metallic properties, etc. These small but significant telescope polarizations are seen in several imaging instruments such as PlanetPol, POLISH, DiPOL, DiPOL2, and HIPPI \citep{2015MNRAS.449.3064B, Hough:2006iz, Bailey:2008fm, Bailey:2010de, Wiktorowicz:2008fm, Berdyugin:2006gn,Berdyugina:2011ca,Berdyugina:2008dj}.  Segmented mirrors, off-axis instrument mounting and data reduction artifacts can all introduce continuum polarization. 

Spectrographs generally are limited in their stability by changes in the optical path (pointing, slit tracking jitter, flexure, dispersive optics sensitivities). Spurious instrumental polarization can also be caused by incomplete scattered light compensation, CCD instabilities, polarization induced by reflective optics (e.g. oblique fold mirrors) and imperfect coatings on optics. Thus, measuring the absolute value of the polarization at high accuracy across the entire continuum of a dispersed spectrum presents challenges to both the instrument and the data analysis pipeline. Many spectrographs have much higher polarization sensitivity across individual spectral lines because the continuum variations can be differentially subtracted across a small wavelength region (cf. \citet{Pereyra:2015gt}) 

In addition to polarization being created by the telescope, the instrument can also scramble or mix incoming polarization states. This mixing, called cross-talk can be from the unpolarized intensity to detected linear or circular polarization or simply mixing between linear and circular states. The LRISp retarders are highly achromatic, reducing the mixing between linear and circular polarization \citep{Keck:2012ub}.  

The LRISp instrument was upgraded to include a second blue camera which also can be used in polarimetric mode \citep{Keck:2012ub,1998SPIE.3355...81M}.  In 2009, the LRIS red detector was upgraded to include higher sensitivity at longer wavelengths \citep{Rockosi:2010ez}.  The atmospheric dispersion corrector (ADC) was mounted in 2007 and includes transmissive prisms \citep{2008SPIE.7014E..53P}. This ADC was tested in 2007 to only marginally impact the measured degree of polarization and angle of polarization for standard stars \citep{Keck:2007wf}.

\section{Observations}

	We observed a range of targets on August 22nd and 23rd 2012.  The 831/8200 grating was used for the red channel at an angle of 37.47$^\circ$ giving coverage from 789nm to 1026nm.  The blue channel used the 300/5000 grism with the 680 dichroic with reasonable sensitivity from 380nm to 776nm though drastic throughput losses were seen long of the 680nm dichroic cutoff.  

	The target list included magnetic stars, brown dwarfs and a range of calibration standards.  EV Lac and V1054 Oph are magnetic flare stars of roughly M3 to M4 type. These stars have roughly known magnetic field strengths and have been studied extensively in the optical and radio \citep{1984ApJ...282..214P, 1994IAUS..154..493S, 1996ApJ...459L..95J, 2000ASPC..198..371J} . We use them here as stars where we expect relatively large and detectable signals.  The star HD20630 is a G5Vv star of BY Dra type.  Though this star is magnetic with known variability, it is listed as a bright unpolarized standard star (in continuum filters) on the UKIRT standard star list \footnote{{\it http://www.jach.hawaii.edu/UKIRT/instruments/irpol/irpol\_stds.html}} and in several publications \citep{1974psns.coll.....G}.  Ceres is the largest main belt asteroid and has a visual magnitude of V=8.

\begin{table*}[!h,!t,!b]
\begin{center}
\caption{\label{Table_LRISp_Observations} Observed targets for LRISp August 22nd and 23rd.}
\begin{tabular}{lcllllll}
\hline
\hline
{\bf Name } 	& {\bf Date}		& {\bf Exp}		&{\bf $\sqrt{N}$} & {\bf Elevation}	&{\bf Defocus?}	&{\bf Spec} 	& {\bf Type}		\\
\hline
\hline
EV Lac		& 22nd			& 20			& 2000	& 64-65			& No			& M4.5V 		& flare star		\\
V 1054 Oph	& 22nd			& 15			& 2000	& 59-60			& No			& M3.5Ve 		& flare star		\\
HD 20630		& 22nd			& 1.2			& 1000	& 73-74			& Yes+		& G5Vv 		& star type BY Dra	\\
2MASS		& 22nd 			& 600		& 350	& 67-84			& No			& L3.5 		& Brown Dwarf		\\
2MASS		& 22nd			& 600		& 350	& 88-77			& No			& L3.5 		& Brown Dwarf		\\
2MASS		& 22nd			& 600		& 350	& 72-59			& No			& L3.5 		& Brown Dwarf		\\
LSRJ		& 22nd			& 600		& 1000	& 74-77			& No			& M8.5V 		& Brown Dwarf		\\
LSRJ		& 22nd			& 600		& 1000	& 74-64			& No			& M8.5V 		& Brown Dwarf		\\
LSRJ		& 22nd			& 600		& 1000	& 62-49			& No			& M8.5V 		& Brown Dwarf		\\
\hline	
Ceres		& 23rd			& 60			& 2000	& 59-54			& Yes-		& Asteroid		& Solar			\\
HD 20630		& 22nd			& 2			& 2000	& 73-74			& Yes		& G5Vv 		& star type BY Dra	\\
HD 174160	& 22nd			& 5			& 1200	& 73-70			& Yes+		& F8V 		& star			\\
2MASS		& 22nd 			& 600		& 200	& 63-77			& No			& L3.5 		& Brown Dwarf		\\
2MASS		& 22nd			& 600		& 200	& 81-88			& No			& L3.5 		& Brown Dwarf		\\
2MASS		& 22nd			& 600		& 200	& 79-64			& No			& L3.5 		& Brown Dwarf		\\
LSRJ		& 22nd			& 600		& 1200	& 77-74			& No			& M8.5V 		& Brown Dwarf		\\
LSRJ		& 22nd			& 600		& 1200	& 73-63			& No			& M8.5V 		& Brown Dwarf		\\
LSRJ		& 22nd			& 600		& 1200	& 59-46			& No			& M8.5V 		& Brown Dwarf		\\
\hline

\end{tabular}	
\end{center}
This table shows all complete polarimetric data sets used for development of this data reduction pipeline. Note that the star we denote as LSRJ is  {\it LSR J18353790+3259545} and the star we denote as 2MASS is an L3.5 brown dwarf:  {\it 2MASS J00361617+1821104}. The star HD 20630 is a magnetic star (BY Dra type) but it is also listed on the UKIRT IRPOL list of unpolarized standards {\it http://www.jach.hawaii.edu/UKIRT/instruments/irpol/irpol\_stds.html} \citep{1974psns.coll.....G}. The statistical upper limit to the signal-to-noise ratio of each polarimetric exposure was computed empirically from polarimetric data. We computed the pixel-to-pixel variance in the polarization spectra after applying high-pass filters to isolate the statistical noise. This noise limit represents the $\sqrt{N}$ limitations from the detected photon flux as an upper limit to the data sensitivity.  We also show the azimuth and elevation range for the telescope to illustrate the variation in local gravity during an exposure. The slit de-rotator was used and set to parallactic for each brown dwarf exposure adding to the gravitational orientation changes between exposures. The telescope was defocused for several bright targets to investigate this technique for increasing the exposure time to saturation.  The Yes- indicates some defocus while Yes+ indicates substantial defocus for very bright targets. The spectral classification and star type from SIMBAD are shown in the last two columns. See the text for details.  \\
\end{table*}

\subsection{Polarimetric Modulation and Demodulation}

In this paper we use the standard Stokes vector formalism to describe polarized spectra.  Linear polarization is denoted as $Q$ and $U$ while circular polarization is $V$.  When we normalize a spectrum by the total intensity, we use lower case symbols.  For instance $q$ = $Q$/$I$. 

We use the general framework for measuring polarization as a {\it modulation} and {\it demodulation} process.  The Stokes parameters are typically described as differences between intensities measured with retarders at different orientations.  In the limit of perfect instrumentation and achromatic optics, a spectropolarimeter can create exposures that mimic the definition of the Stokes parameters. Several modulation strategies are in use in solar, space and night time applications in order to balance the need for efficiency, redundancy, error checking through {\it null spectra} and for simplicity of data analysis \citep{Tinbergen:2005up, 2003isp..book.....D, delToroIniesta:2000cg, 2013pss2.book..175S, Nagaraju:2007tn, Tomczyk:2010wta, Snik:2012jw, Snik:2009va, deWijn:2010fh}. In the typical notation, the instrument modulates the incoming polarization information in to a series of measured intensities (${\bf I}_{i}$) for $i$ independent observations via the modulation matrix (${\bf O}_{ij}$) for $j$ input Stokes parameters (${\bf S}_j$): 

\begin{equation}
{\bf I}_{i} = {\bf O}_{ij} {\bf S}_{j}
\end{equation}

In most night-time polarimeters, instruments and associated data analysis packages a modulation matrix that separates and measures individual parameters of the Stokes vector as well as providing redundant information for use in characterizing instrument performance \citep{1993A&A...278..231S, Donati:1997wj}.  In the {\it Stokes definition} modulation scheme, there are 6 exposures recorded each corresponding to an independent Stokes parameter ($QUV$).  

\begin{equation}
\label{normmod}
{\bf O}_{ij} =
 \left ( \begin{array}{rrrr}
 1   	& +1	&  0	&  0	\\
 1 	&  -1	&  0 	&  0	\\
 1 	&  0 	& +1	& 0	\\
 1 	&  0	& -1	&  0	\\ 
 1 	&  0	& 0	&  +1	\\ 
 1 	&  0	& 0	&  -1	\\ 
 \end{array} \right ) 
\end{equation}

In ESPaDOnS, FORS and other instruments, additional redundancy is achieved by making another set of measurements using the same modulation matrix but with all retarders rotated by 180 degrees \citep{Bagnulo:2009bz, Donati:1999dh, 1993A&A...278..231S}.   In LRISp, this type of modulation is accomplished in two separate optical configurations (for two separate exposures). A rotating half-wave super achromatic retarder plate (HWP) is mounted in front of the analyzer. The HWP is rotated in a sequence of [0$^\circ$, 45$^\circ$, 22.5$^\circ$, 67.5$^\circ$] in order to accomplish linear polarization modulation in 4 exposures with beam swapping.  Circular polarization is measured by rotating a second quarter-wave achromatic retarder plate (QWP) into the optical path using the calibration filter wheel.  This QWP is fixed in a single orientation and modulation is accomplished by rotating the HWP by 0$^\circ$ and 45$^\circ$ behind the QWP.  

In the Stokes definition scheme, calculation of each Stokes parameter from intensity spectra follows Equation \ref{eqn_stokesdef_mod} is implemented as a series of normalized intensity differences recorded in two exposures assuming perfect modulation and achromatic optics:

\begin{equation}
\label{eqn_stokesdef_mod}
q = \frac{Q}{I} = q_0 + q_1 =  \frac{ I_0 - I_1}{I_0+I_1} - \frac{I_2-I_3}{I_2+I_3}
\end{equation}

	We wish to highlight that these normalized intensity differences when assumed to represent a Stokes parameter can introduce several types of instrumental errors while also ignoring cross-talk. In dual beam systems, two pairs of spectra are recorded in two exposures. Thus each part of the ratio is subject to instrumental uncertainties that are introduced between exposures. 

	In many night time spectropolarimeters, the instrument is designed so the cross-talk is below some nominal design value. Chromatic effects are often just minimized and subsequently left uncalibrated, but no additional spectra are used to compute a Stokes parameter. By using additional spectra, or additional calibrations, cross-talk can be further minimized. Each source of polarimetric error must be considered when choosing an optimal modulation scheme and associated data processing algorithms. 

	Other instruments choose to modulate and measure all Stokes parameters using only four measurements or pursue less redundant but more efficient schemes \citep{deWijn:2010fh, delToroIniesta:2000cg, Tomczyk:2010wta, Snik:2012jw, Snik:2009va, Nagaraju:2007tn, Keil:2011wj, Elmore:2010ip, 1992SPIE.1746...22E}. For instance, one can use alternate retardance and fast axis orientations to give a modulation matrix that uses only four exposures to measure a Stokes vector with maximal efficiency. 

One recovers the input Stokes vector from the series of intensity measurements by inverting the modulation matrix (${\bf O}$). If the matrix is not square, and non-unitary (as in the Stokes definition scheme) one can simply solve the over-specified system of equations via the normal least squares formalism: 

\begin{equation}
\label{eqn_demod}
{\bf S} = \frac{ {\bf O}^T {\bf I} } { {\bf O}^T {\bf O}}
\end{equation}

Other modulation schemes are easily crated using tunable liquid crystals and the modulation matrix does not need to have any particular symmetry.  We have implemented this ourselves on other spectropolarimeters \citep{Harrington:2010km,Harrington:2011fz}. In systems with more complex or less redundant modulation schemes, additional calibrations with upstream polarizers and retarders are often used to achieve the highest calibration accuracy and remove residual chromatic effects from the demodulation process. The demodulation process of Equation \ref{eqn_demod} can be used regardless of modulation scheme.  

Though often not performed, these same kind of demodulation processes can be applied to {\it Stokes definition} type modulation schemes to remove residual chromatic errors if the imperfections are estimated through a calibration procedure. For instance, in LRISp, there are calibration polarizers mounted in the filter wheel ahead of the rotating HWP retarder. This polarizer can be used to measure some of the chromatic properties of the HWP and to modify the modulation matrix of Equation \ref{normmod} to account for imperfections.  Polarized standard stars or more elaborate calibration optics can be used to derive the system Mueller matrix to correct for some residual cross-talk.  Alternatively, the daytime sky calibrations we outline here and elsewhere can be used to measure the system Mueller matrix and to apply corrections to the demodulated spectra to account for any uncalibrated cross-talk to some residual error levels \citep{Harrington:2011fz}, Harrington et al. 2015.

\subsection{Geometric calibrations - wavelength stabilization}

\begin{figure} [!h, !t, !b]
\begin{center}
\includegraphics[width=0.75\linewidth, angle=90]{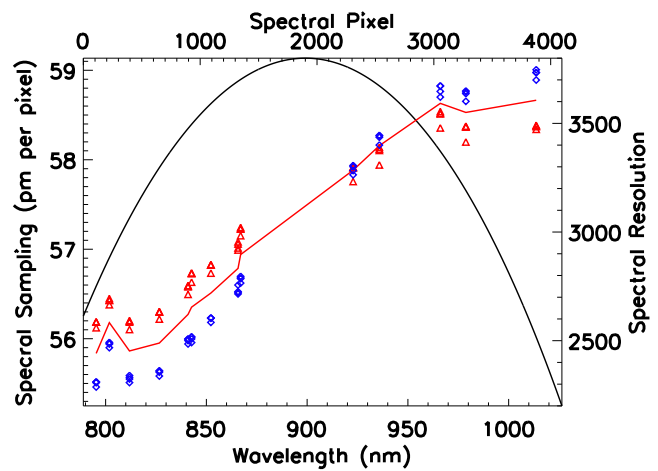}
\caption{ \label{waveln_sampres} The derived spectral sampling and spectral resolution for the LRISp red channel. The black parabolic curve shows the spectral sampling (pico meters per pixel) using the left hand y axis. The mapping between the wavelength solution and each spectral pixel is found by comparing the top and bottom x axes.  The spectral resolution is derived as the FWHM of the arc lamp Gaussian fits as defined in the text. The arc lines typically have a FWHM of 4.5 to 5.5 spectral pixels. The spectral resolution is derived as the wavelength decided by the FWHM of the arc line fits and is shown with the symbols using the right hand y axis. The resolution was between 2500 and 3500 across the sampled wavelengths. The blue symbols show the spectral resolution of the top polarized beam.  The red symbols show the spectral resolution of bottom polarizers beam.  The red curve in between these symbols shows the average spectral resolution of both top and bottom beams.   }
\end{center}
\end{figure}

	The first steps in spectral extraction is locating the spectral orders and identifying the basic optical configuration.  When combining 6 exposures to make a single $quv$ spectral data set, any instrumental instability can produce spurious signals. In general, spectral order curvature, anamorphic magnification, and tilt of the slit image against the ccd pixel grid all can impact polarimetric data. This is especially true given telescope guiding imperfections and instrument flexure. For the LRISp red channel, the basic parameters of our optical extraction are shown in Figure \ref{waveln_sampres}.  We derive several parameters from fits to several arc lamp calibration exposures. The spectral resolution is derived from arc lamp calibration exposures.  The resolving power (R=$\lambda$ / $\delta\lambda$) shown in Figure \ref{waveln_sampres} is R=2500 at 800nm rising to R=3500 at 1000nm. In this configuration, the spectra are oversampled. From Gaussian fits to arc lamp spectral lines, we find the spectra to be sampled at 4.5 pixels to 5.5 pixels in the Gaussian full-width half-max (FWHM). With this over sampling of 0.56{\AA} to 0.59{\AA} per pixel, we can test for several instrumental artifacts and apply several types of data post processing filters to remove noise sources.

	With the arc line exposures, we measure how the wavelength coordinates change as the HWP rotates through the typical modulation sequence of 0$^\circ$, 45$^\circ$, 22.5$^\circ$, 67.5$^\circ$. This rotating HWP causes a drift of roughly 0.15 pixels between the two separate modulation states. For the LRISp HWP as mounted, the offsets average [0, 0.08, 0.13, 0.10] pixels referenced to the first exposure. There is also a mild wavelength dependence across the CCD. Note that the two Stokes $q$ exposures would show a wavelength drift of 0.1 pixels in between modulated images resulting in imperfect subtraction introducing artifacts resembling the derivative of the intensity profile with wavelength.  However, the Stokes $u$ exposures would show substantially less wavelength drift between modulation states, but would be offset from Stokes $q$ spectra by 0.1 pixels in wavelength.

	Slit guiding for Keck is software-referenced and the user can vary the stellar location along the length of the slit. In addition, we describe later how we had substantial guiding drifts tracking our targets. We observed at higher elevations and see expected drifts for an altitude-azimuth telescope with a low-bandwidth guider control system.  With drift of the optical beam along the slit, uncorrected geometrical tilt of the dispersed spectra will lead to wavelength drifts between exposures.

	For the red channel, we find significant tilt of the monochromatic slit images against the pixel rows. There is roughly 2 pixels of wavelength change on the CCD pixels from the bottom to the top of the imaged slit over the 300 spatial pixels sampled.  We observed up to 40 spatial pixels of guiding drift during our observing run. This spatial drift combined with the spectral tilt would give wavelength instabilities of up to half a pixel if this geometrical effect is not compensated in the pipeline.

\begin{figure} [!h, !t, !b]
\begin{center}
\includegraphics[width=0.75\linewidth, angle=90]{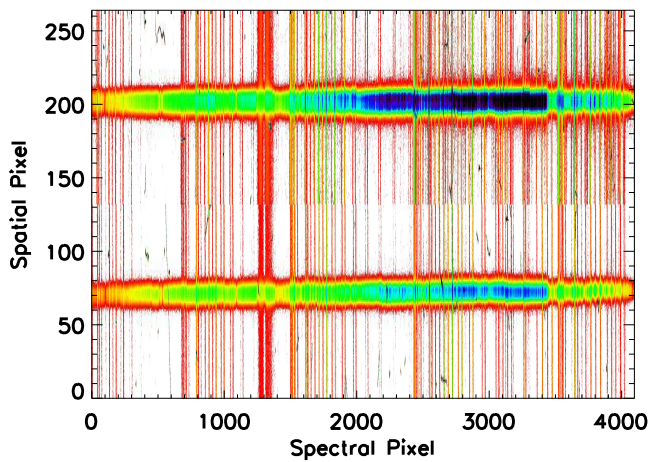}
\caption{ \label{2d_example} An example of an extracted 2-D spectrum for a typical LSRJ brown dwarf observation. The two beams are shown after tilt-correction as observed August 22nd after geometric extraction. Several night sky glow lines are visible as the vertical stripes.  The top beam has noticeably higher throughput than the bottom beam as was typical for this spectrograph configuration. The intensity was linearly scaled with blue and black colors corresponding to the highest intensities.}
\end{center}
\end{figure}

To compensate for this geometrical tilt, the data is linearly up-sampled to a 0.01 pixel grid and then shifted spectrally to compensate for the tilt.  The data is then averaged back down to nominal 1-pixel sampling.  An example of an extraction after tilt correction is shown in Figure \ref{2d_example}.  In this Figure, the stellar spectrum is in the center of the two extracted beams.  The 4000 spectral pixels of the detector are shown on the X-axis.  We only show a subset of the extracted spatial pixels to clearly show the stellar spectrum.

\section{Automatically Removing Cosmic Rays With Iterative Filters}

\begin{figure} [!h, !t, !b]
\begin{center}
\includegraphics[width=0.75\linewidth, angle=90]{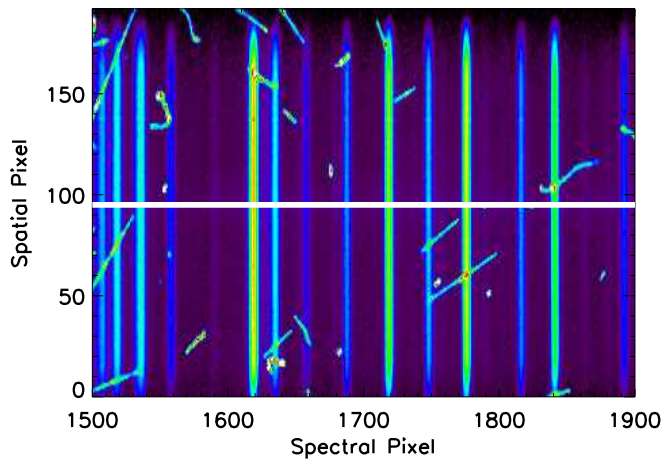}
\caption{ \label{sky_background2d} This Figure shows a small region of the 2-D spectrum used to extract and remove the sky-glow lines in a single polarimetric exposure. The two polarized beams are shown separated by a white line. Typical cosmic ray hit rates are seen. }
\end{center}
\end{figure}

The LRIS red channel pixels are 300 microns deep, giving a fairly high rate of cosmic ray hits in long exposures \citep{Rockosi:2010ez}.  An example of the background region in a typical 10 minute brown dwarf exposure is shown in Figure \ref{sky_background2d}.  Bright night sky glow lines are seen in addition to many cosmic ray hits.  Cosmic ray damage in this CCD often spans 10s of pixels and most spectral pixels are contaminated at some level, making sensitive polarimetry difficult. 

\begin{figure} [!h, !t, !b]
\begin{center}
\includegraphics[width=0.75\linewidth, angle=90]{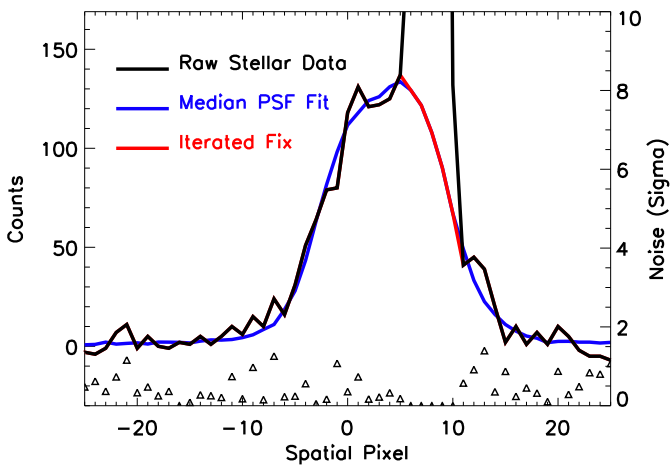}
\caption{ \label{cosray_rejection_example} An example of the cosmic ray iterative filtering process. The black curve shows the raw stellar spatial profile (a cut through the data orthogonal to the spectral direction representing the local seeing, telescope jitter and optical imperfections). A very large cosmic ray hit is seen damaging several pixels in the middle of the spatial profile at pixels 5 to 10. The wide spread damage requires the iterative solution to reject the damaged data. After iterating, 5 spatial pixels are rejected and ignored in the shift-n-scale profile fit. The blue curve shows the median spatial profile used in the fit.  The red curve shows the spatial profile replacements. The right hand y axis shows the noise estimates used in the filter.  The triangle symbols below show the noise estimates on a per pixel basis used in the rejection. A 2-sigma filter was applied in this case. }
\end{center}
\end{figure}

We define the {\it spatial profile} as a trace through the data that is orthogonal to the wavelength direction after geometric calibration of the extracted spectral data has been performed.  This spatial profile contains the atmospheric seeing, telescope jitter, guiding imperfections, optical imperfections (ghosts) and can be used to apply data-derived filters for various error sources.  An example spatial profile is shown in Figure \ref{cosray_rejection_example}.  

An iterative method has been developed to use the spatial profile to filter out these cosmic ray hits based on what's called {\it optimal extraction} \citep{Horne:1986bg, Marsh:1989jo}. The spatial profile is computed over a range of wavelengths. We find that 100 spectral pixels is a good compromise between increasing the SNR of the spatial profile and ensuring that the wavelength of the spatial profile matches that of the wavelength to be filtered.   

The first step in the filter is to shift and scale the median spatial profile to the individual spatial profile of interest. The least squares solution implemented over the $i$ spatial pixels that independently contribute to the problem. 

\begin{figure*} [!h, !t, !b]
\begin{center}
\includegraphics[width=0.65\linewidth, angle=90]{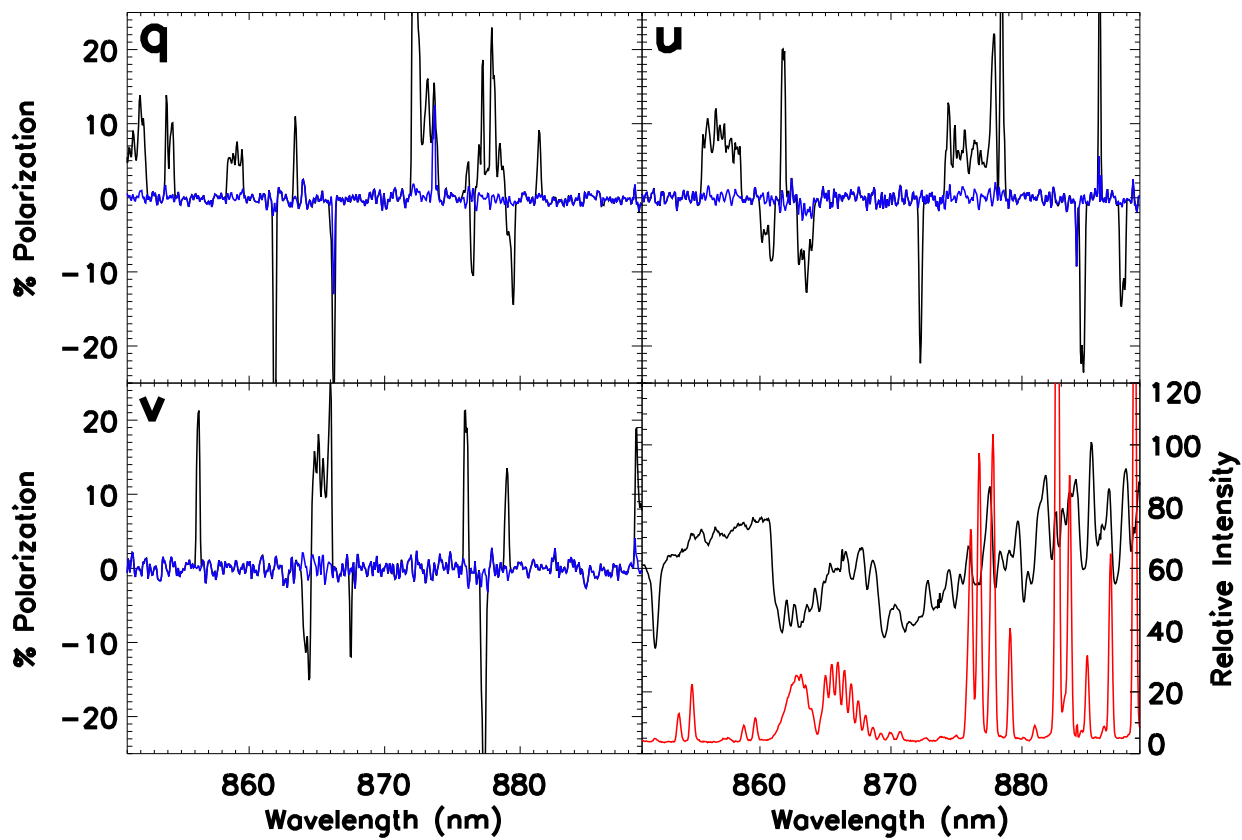}
\caption{ \label{polarimetry_2mass} This Figure shows 2MASS set 3 from August 23rd. Stokes $quv$ polarization was computed using the standard {\it Stokes definition} modulation assumption with beam swapping as $\frac{ I_0 - I_1}{I_0+I_1} - \frac{I_2-I_3}{I_2+I_3}$. The cosmic ray rejection filter was applied with the noise threshold of 1.8$\sigma$. A 10-count background noise level was set. A 50-pixel smoothing width was used to compute the median spatial profile in the in the shift-n-scale fitting algorithm. The black lines show the computed polarization spectra before cosmic ray filtering.  The blue line shows the resulting polarization after the iterative cosmic ray filtering is applied. Cosmic rays are present in a substantial fraction of each $quv$ spectrum. The lower right panel shows the intensity spectrum with a black line in units of 100s of detected counts per spectrum. The sky glow spectrum extracted from spatial pixels outside the region illuminated by the star is shown in red. For 2MASS, the sky glow lines often are brighter than the star at certain wavelengths. }
\end{center}
\end{figure*}

We model the shift-and-scale problem with a general notation where $D$ denotes the data to be fit and $P$ denotes the profile used to do the fitting. The derivative in the spatial direction is $\partial x$ The data ($D$) is modeled as a sum of three terms:  a constant (a) times the profile ($P$), a constant (b) times the derivative of the profile, and an additive constant (c). 

\begin{equation}
D = a P + b \frac{\partial P}{\partial x} + c
\end{equation}

The total error to be minimized is a sum over all spatial pixels: 

\begin{equation}
\label{eqn_rotmat_error}
E = \sum \limits_{i}  \epsilon^2  = \sum \limits_{i} (D - a P - b \frac{\partial P}{\partial x} - c)^2
\end{equation}

In order to solve the least squares problem, we take the partial derivatives with respect to the three coefficients (a,b,c) and set them equal to zero:

\begin{equation}
\label{partials}
\frac{ \partial E } {\partial a} = \frac{ \partial E } {\partial b} = \frac{ \partial E } {\partial c} = 0
\end{equation}

After doing the partial derivatives, collecting terms and solving for 0, we get a system of three equations for three variables:

\begin{equation}
\left ( \begin{array}{r}
PD \\
\frac{\partial P}{\partial x} D \\
D \\
\end{array}  \right ) = 
\left ( \begin{array}{lll}
P^2						& P \frac{\partial P}{\partial x }		& P						\\
P \frac{\partial P}{\partial x} D	&     \frac{\partial P}{\partial x }^2	& \frac{\partial P}{\partial x} D	\\
P						&     \frac{\partial P}{\partial x }		& 1						\\
\end{array}  \right )
\left ( \begin{array}{lll}
a \\
b \\
c \\
\end{array} \right ) 
\end{equation}

With the least-squares solution, we can compute the difference between the shift-n-scaled median spatial profile and the individual spatial profile to be corrected.  The difference between these two spatial profiles is then compared against a measure of the expected noise at every spatial location.  The expected noise has two contributing terms.  One term is a constant background noise estimate representing the read, dark and scattered light background variation.  The second term accounts for the spatially varying flux levels and is proportional to the square root of the number of detected counts.   This shot-noise term is computed using the shift-n-scaled median spatial profile. 

This process is iterated until convergence is achieved and no spatial pixels show noise levels above a user-determined set of thresholds.  At each step of the loop, the single worst offending point above the noise threshold is rejected from consideration by setting the weighting in the fit to zero.  Once the iteration has converged, it delivers a shift-n-scaled median profile that fits all non-rejected points below the noise threshold.  After convergence, all rejected spatial profile points are replaced with the shift-n-scaled median spatial profile values.  This iteration ensures that very large cosmic ray strikes are properly corrected without undue influence in the fitting process.

\begin{figure*} [!h, !t, !b]
\begin{center}
\includegraphics[width=0.35\linewidth, angle=90]{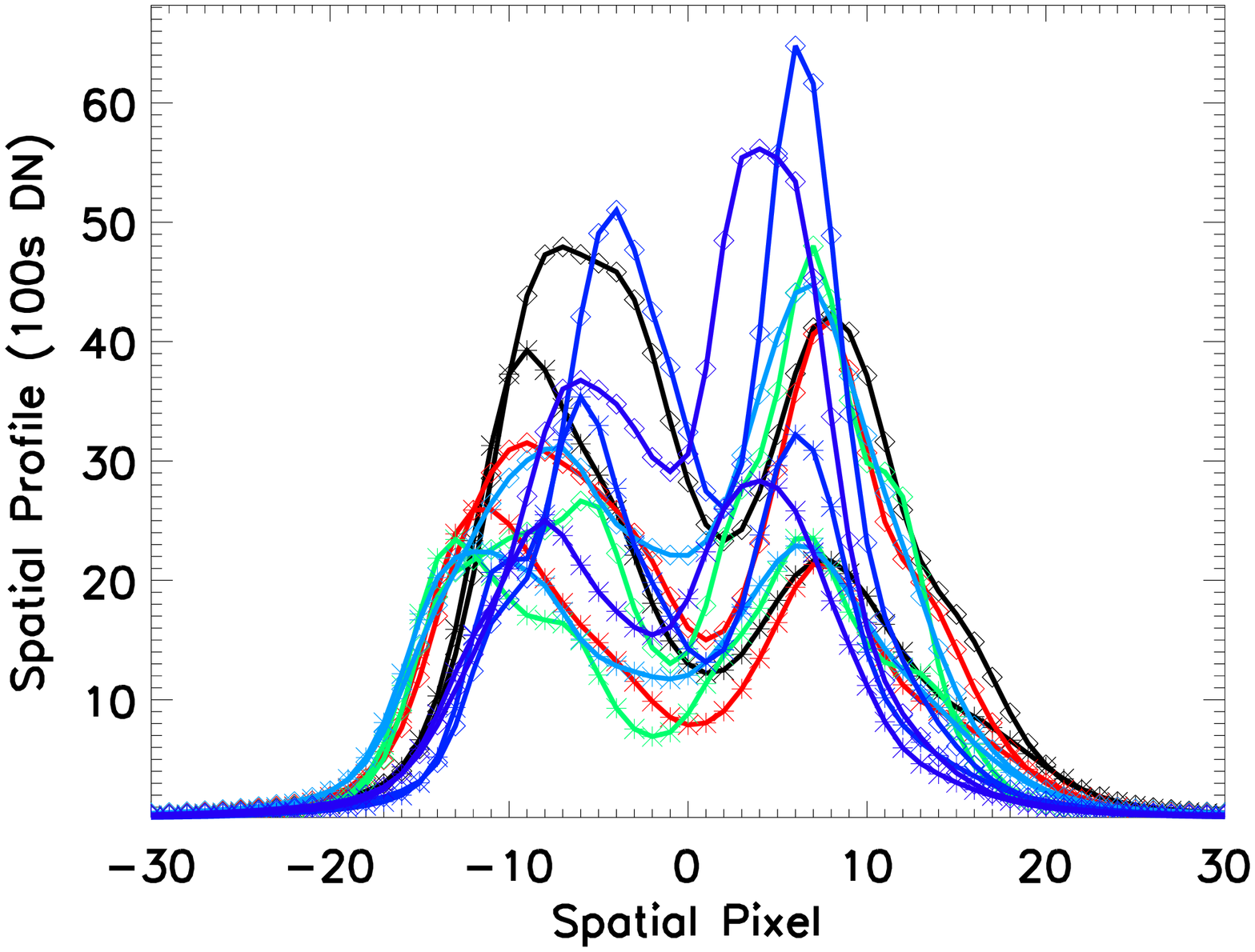}
\includegraphics[width=0.35\linewidth, angle=90]{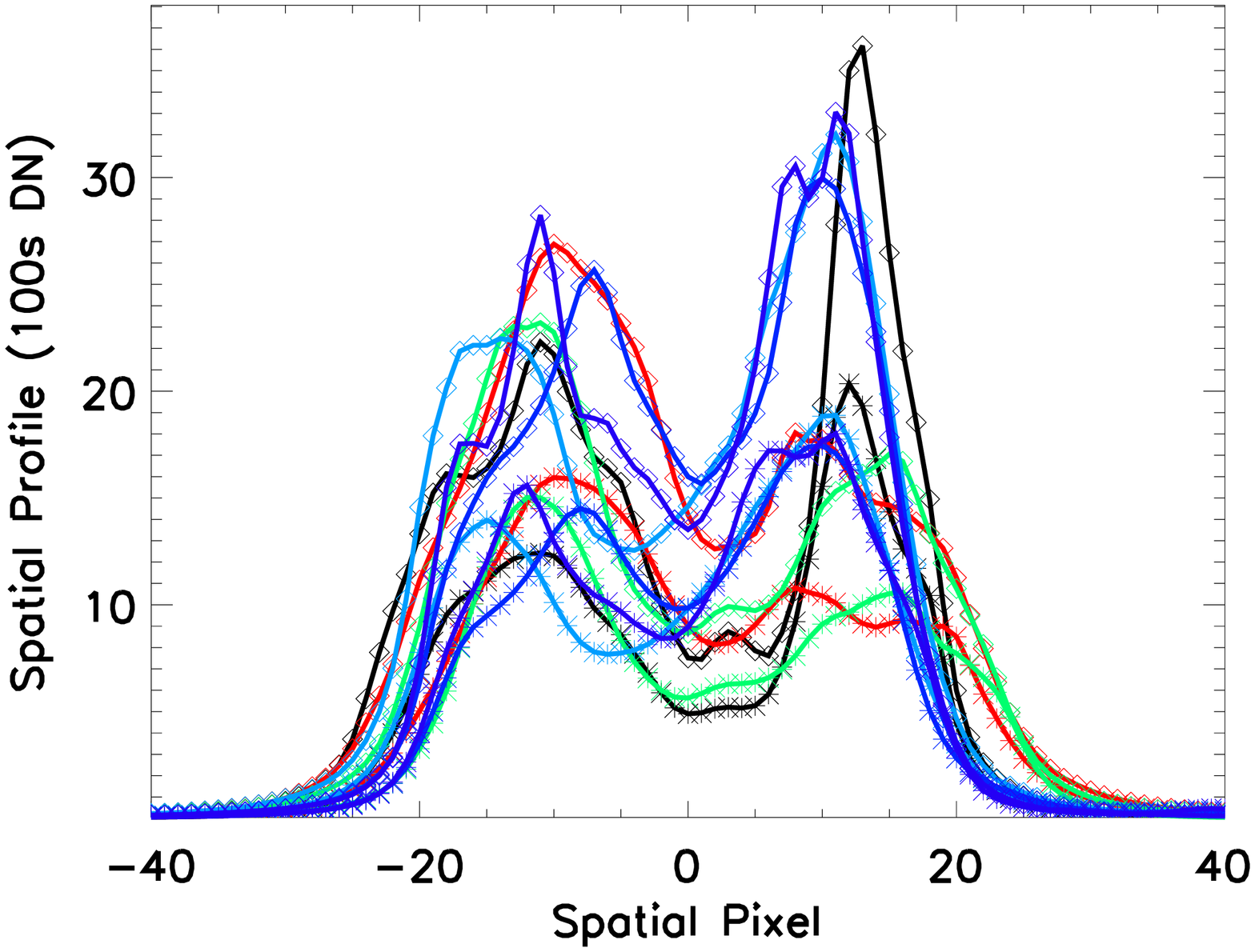}
\includegraphics[width=0.35\linewidth, angle=90]{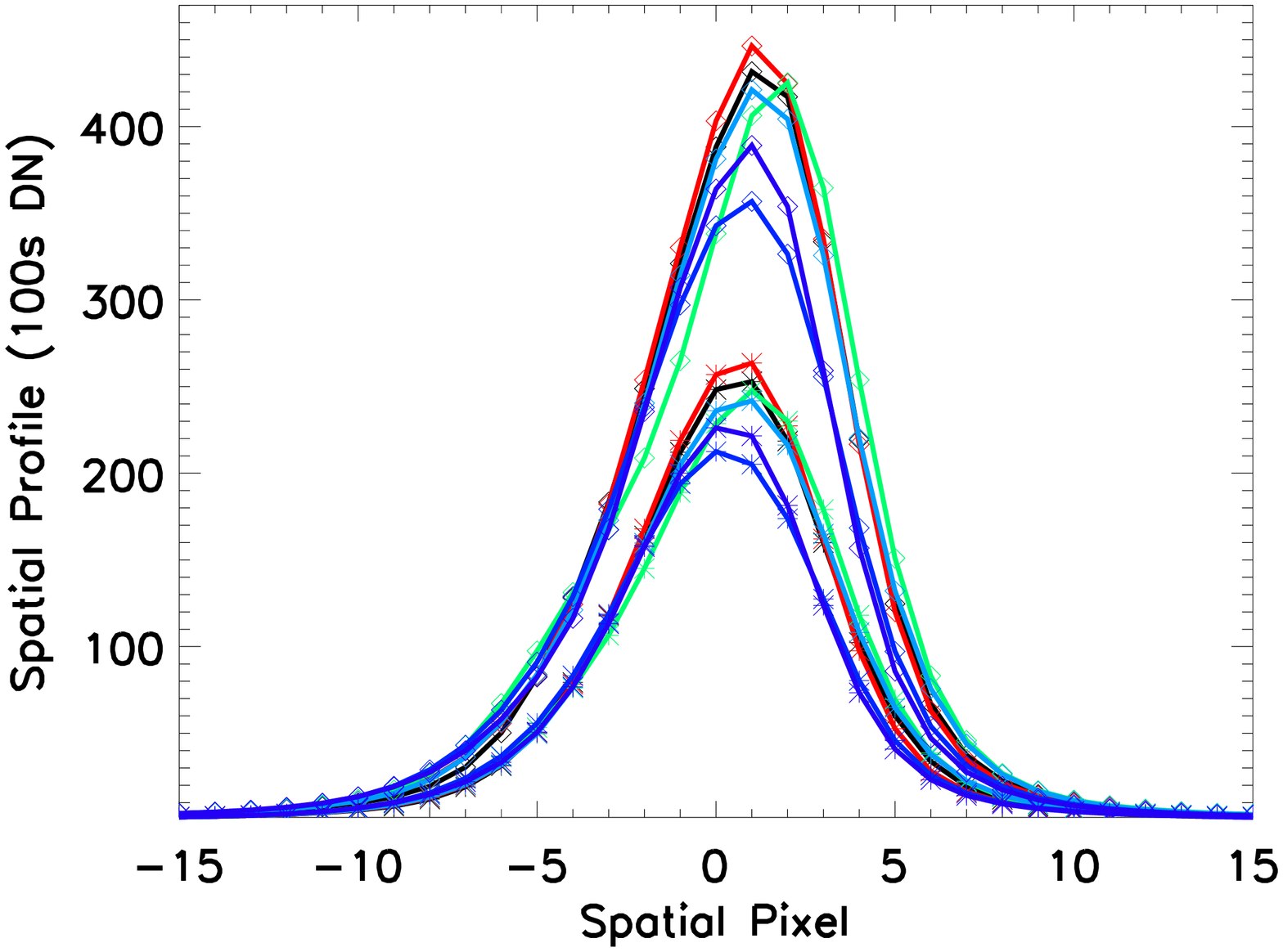}
\includegraphics[width=0.35\linewidth, angle=90]{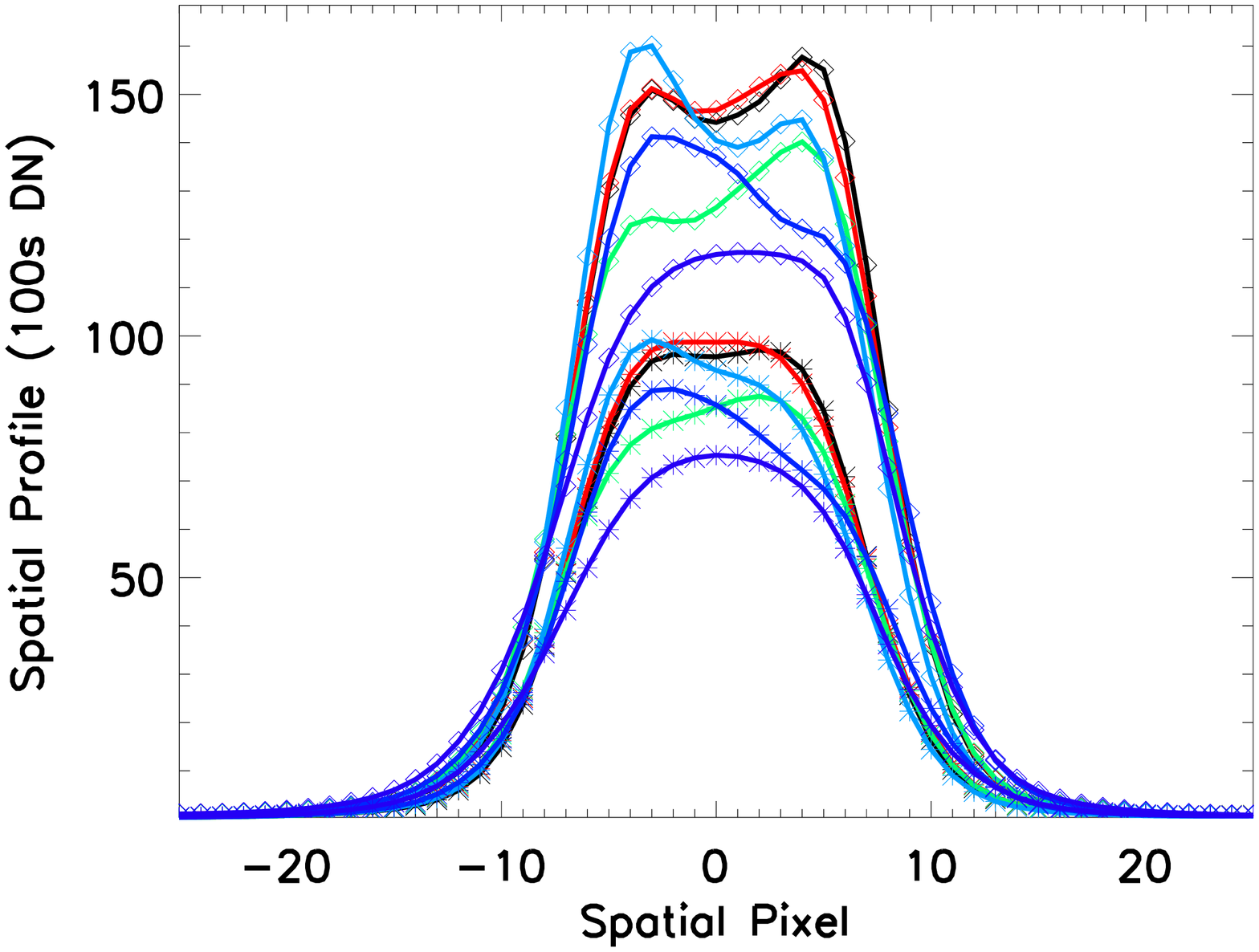}
\caption{ \label{psfs_polarimetric_manystars} The spatial profile of the extracted spectra after spatial trimming for guider correction and optimal extraction filtering. The unpolarized standard HD 174160 is in the upper left from August 23rd with an extraction width of 60.  The unpolarized but magnetic BY Dra star HD 20630 is in the upper right from August 22nd with an extraction width of 80.  The magnetic flare star EV Lac is in the lower left from August 22nd with an extraction width of 30.  The asteroid Ceres is in the lower right from August 23rd with an extraction width of 50. All 12 spatial profiles are shown (6 exposures for $quv$ and 2 polarized beams per exposure). }
\end{center}
\end{figure*}

An example of this process is shown in Figure \ref{cosray_rejection_example}.   In this example, a spatial profile for a 2MASS brown dwarf exposure on August 22nd is shown in the black curve.  A very large cosmic ray hit is seen at spatial pixels 5 to 10. This cosmic ray shows a count level an order of magnitude larger than the detected stellar flux.  The filter iterates through the shift-n-scale process and identifies spatial pixels with noise limits above the user-defined thresholds for read and shot noise. After all points have been rejected, the red curve shows the values used to repair the spatial profile.  The triangle symbols in Figure \ref{cosray_rejection_example} show the residual noise levels for the points included in the fit.  The user specified threshold was set at 2$\sigma$ for this example. 

Typical performance of the cosmic ray filter is shown in Figure \ref{polarimetry_2mass}.  The black curve shows the $quv$ spectra with substantial cosmic ray hits contaminating a large fraction of the wavelengths covered.  The blue curve shows the corresponding filtered $quv$ spectra with effective removal of the cosmic ray hits.

\section{Spatial profiles and slit guider tracking}

The Keck slit guider was used to acquire and track our targets.  In some cases the guiding delivered spatial profiles which remained centered to within 5 spatial pixels for the duration of a 6-exposure polarimetric data set.  For other exposures, there was drift of over 50 spatial pixels. This corresponds to roughly 20\% of the slit length (290 spatial pixels in our extractions).  Thus our pipeline was configured to compute the center of light for each exposure of a polarimetric data set.  The data was extracted around this detected center-of-light for each exposure to compensate for the guiding drift. 

A common technique to increase the exposure time until saturation and increase the duty-cycle of measurements is to defocus the telescope. We tested the effects of defocus on achieving high SNRs without saturation on very bright targets. During spectral extraction, the spatial profile width varied from roughly 10 pixels full width at 10\% max for in-focus brown dwarf targets. For the bright unpolarized standards, this spatial width increased to over 40 pixels at 10\% max. This focus shift changes the optical beam footprint through both QWP and HWP retarders, the polarizing beam splitter and all downstream spectrograph optics and care must be taken with calibration (cf. \citet{Tinbergen:2007fd}).  

Figure \ref{psfs_polarimetric_manystars} shows the spatial profiles for four stars to illustrate the change in telescope focus. The top panels show highly defocused observations of bright stars.  The bottom panels show an in-focus brown dwarf on the bottom left and a mildly defocused Ceres exposure on the bottom right. The data reduction pipeline allows for a variable spatial extraction width that changed between 30 and 80 pixels to accommodate this wide ranging defocus.  In addition, by including the minimal number of spatial pixels needed to capture the delivered target flux, we minimize the impact of cosmic rays and other detector noise contributions (cosmetics, bad columns, etc).

\section{Flexure Compensation}

There is substantial wavelength drift even within a polarimetric data set due to instrument flexure. These wavelength instabilities cause a major source of spurious instrumental polarization.  Measurement, compensation and accurate error budgeting of these types of systematic effects are critical to interpretation of spectropolarimetric results (c.f. \citet{2013A&A...559A.103B}). Some instruments use a more redundant modulation scheme where there are 4 exposures per Stokes parameter (using the Stokes definition modulation scheme).  These redundant schemes do provide for error assessment through the so-called {\it null-spectrum} in addition to further removal of some instrumental error sources.  However, requiring 4 exposures also introduces the possibility of wavelength jitter over long exposure times with gravity changes (for Cassegrain instruments like LRISp).  \citet{2013A&A...559A.103B} also found that using 4 dual-beam exposures (instead of 2 in our scheme) introduced more systematic errors through flexure instabilities.  Additionally, some of our science targets have fast rotation periods and there is noticeable change over even the 2 polarimetric exposures of a single Stokes parameter spectrum.  Thus any science campaign must balance the error budgeting between the different types of systematic errors based on the specific use case (exposure time, flexure predictions, null spectrum requirements, overly redundant modulation for error suppression, etc). 

\begin{figure} [!h, !t, !b]
\begin{center}
\includegraphics[width=0.95\linewidth, angle=0]{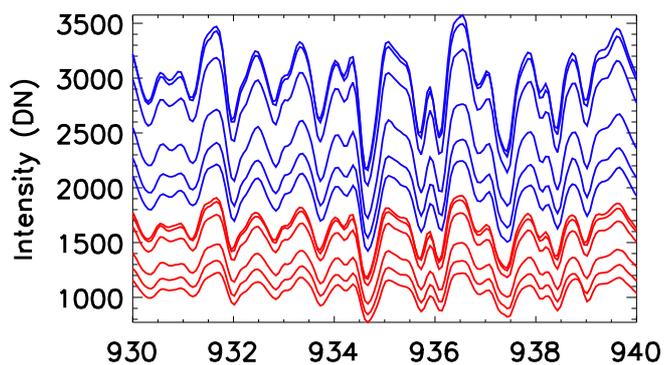}
\caption{ \label{lsrj_telluric_spectrum} The intensity spectra for the LSRJ data set 1 from August 22nd. The detected flux is shown for each of the 6 exposures and 2 polarization states (dual beam). The polarization state 0 beam is red while polarization state 1 is blue.}
\end{center}
\end{figure}

In the brown dwarf data sets, a polarimetric observation can last over an hour. For faint targets, we can use sky glow lines to clearly align the wavelengths for each portion of the spectrum with a sufficient number of lines. However, in brighter targets with shorter exposure times, we must use the telluric absorption lines as an absolute wavelength reference.  The 930nm to 940nm bandpass has many absorption lines that can easily be used for correlation.  Figure \ref{lsrj_telluric_spectrum} shows the telluric wavelength region for the first data set recorded on August 22nd. The variation in detected intensity is due mainly to telescope guiding drifts. The two different orthogonally polarized beams produced by the polarizing beam splitter have noticeable throughput variations. The red curves from polarization state 0 are roughly half the flux of the polarization state 1 beams.

We derive a wavelength correction for each exposure by running a cross correlation analysis. All spectra are up-sampled by a factor of 100 to derive correlations in 0.01 pixel bins.  The polarization state 0 beam of the first exposure is used as a reference for all subsequent exposures and polarization states in the 6 exposure set. From these correlation functions we find the peak and derive a wavelength shift to align all exposures. These telluric corrections were typically less than a pixel but with varying behavior with exposures and between the dual orthogonally polarized beams. 

Flexure that is common to both dual orthogonally polarized beams will be removed to first order with the beam swapping applied in the nominal {\it Stokes definition} modulation scheme. For the Ceres observations we use as a fringe standard, we find a 0.5 pixel drift in wavelength but this shift is consistent between the two polarization states. Provided the wavelength drift between the two polarized beams is consistent, the primary error proportional to intensity derivatives will be canceled when calculating the polarization spectra using the difference method (a-b)/(a+b). Additional errors proportional to the second derivative of the intensity will be come significant if the drift between exposures is large and uncompensated. 

\begin{figure} [!h, !t, !b]
\begin{center}
\includegraphics[width=0.95\linewidth, angle=0]{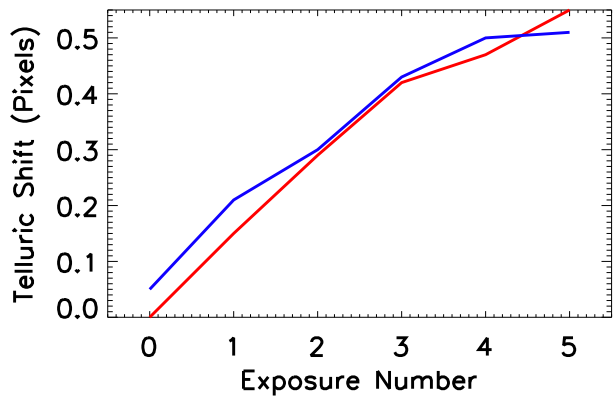}
\caption{ \label{lsrj_telluric_pixel_shifts_set2} The pixel shifts for the LSRJ data set 2 from August 22nd. The polarization state 0 beam is red while polarization state 1 is blue. }
\end{center}
\end{figure}

Even though these wavelength shifts are small, they can have large impact on the computed polarization if differential effects occur. For some of our data sets, we measure wavelength offsets of roughly 0.2 pixels between the dual orthogonally polarized beams. As an example, the derived spectral pixel shifts for the second August 22nd LSRJ data set is shown in Figure \ref{lsrj_telluric_pixel_shifts_set2}. There is a noticeable difference in behavior between the polarized beams for the final exposure 5.

This difference of 0.2 pixels in wavelength for one exposure, though small, introduces a very large systematic error in polarization. Since the computed $quv$ profiles are differences in intensities, any wavelength drift imprints a $quv$ signature that is proportional to the derivative of the intensity with wavelength. This small 0.2 pixel wavelength drift is enough to imprint a 0.5\% Stokes $v$ signature change when the corresponding intensity spectra has a strong absorption line. Applying a wavelength drift correction is critical to deriving accurate polarization spectra with LRISp. 

\begin{figure} [!h, !t, !b]
\begin{center}
\includegraphics[width=0.95\linewidth, angle=0]{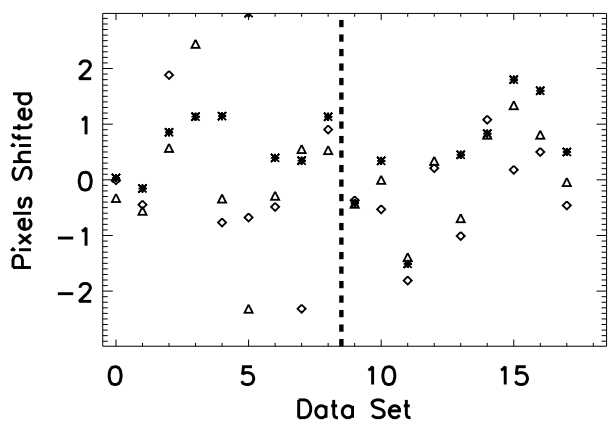}
\caption{ \label{ripple_spectral_drift_all_stars} The wavelength drift (in pixels) for all $quv$ data sets derived through cross correlation of telluric lines with the nominal Ceres data set on August 23rd. The symbol for $q$ is the triangle, $u$ is the diamond and $v$ is the cross. The left side shows August 22nd data while the right side shows August 23rd.  The wavelength drift is roughly $\pm$ 2 pixels.}
\end{center}
\end{figure}

Another added benefit of including cross-correlation analysis in our LRISp pipeline is that all spectra can be easily referenced to a common wavelength.  Figure \ref{ripple_spectral_drift_all_stars} shows the wavelength shift (in pixels) derived from correlating the Ceres telluric spectra with all exposures for all August 22nd and 23rd data sets.  The flexure derived roughly $\pm$2 pixels within a night and a fraction of a pixel within a data set. 

Additionally, we have tested the wavelength stability across the detector. In principal, the flexure could induce wavelength changes that are not completely removed by a single shift for all wavelengths.  Any optical distortion or second order effects could create a more complex functional dependence on the wavelength solution with flexure.  We ran a cross-correlation of the intensity spectra within a complete polarimetric data set. Essentially we find that the wavelength solution is well corrected by a single shift of all spectral pixels to within at least 0.05 pixels.  We find perturbations in wavelength regions where we detect spectropolarimetric signatures, as we would expect from magnetic field effects.  As expected, with the highly defocused standard stars without the use of the ADC, there are shifts in the wavelength solution of up to 1 pixel across the detector.  However, this drift only influences the unpolarized standard star observations.

\section{Spectral Fringes}

Another feature of LRISp data at high SNR is a spectral fringes caused by interference within the achromatic retarders.  Fringes such as these can be common in night-time spectropolarimeters such as the  Intermediate dispersion Spectrograph and Imaging System (ISIS) on the William Herschel Telescope (WHT) or the Anglo-Australian Telescope (AAT) with the Royal Greenwich Observatory (RGO) spectrograph  \citep{Aitken:2001ih, Harries:1996vf, Donati:1999dh}.  For most of these spectrographs over relatively limited wavelength regions, the spectropolarimetric fringe follows a simple functional form. Chirp functions or simple harmonic filters in Fourier space are enough to suppress the fringes below the statistical noise limits.  

\begin{figure} [!h, !t, !b]
\begin{center}
\includegraphics[width=0.75\linewidth, angle=90]{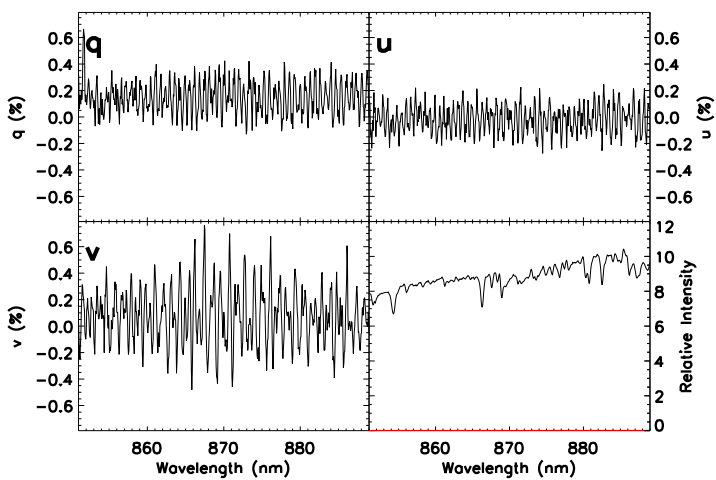}
\caption{ \label{evlac_spectral_ripple} This Figure shows the $quv$ spectra in the 850nm region from EV Lac. The SNR is estimated to be about 1200 to 1800 in each Stokes parameter.  A strong spectral fringe is seen dominating the $quv$ spectra. The fringe amplitude is well above the statistical noise limits and is well sampled with many spectral pixels.}
\end{center}
\end{figure}

For LRISp, the fringe depends on the Stokes parameter being measured. Stokes $qu$ measurements use only a single rotating half-wave retarder (HWP) giving a fringe at 0.2\% amplitude at 850nm.  Measuring Stokes $v$ with LRISp involves adding a second achromatic quarter wave plate (QWP) fixed in front of the HWP.  This second optic complicates the fringe and increases the amplitude to roughly 0.5\% at 850nm.  Figure \ref{evlac_spectral_ripple}.

The exact form of the fringe power spectra depend on several things.  First, the telescope was defocused for the brighter targets.  During spectral extraction, the spatial profile width varied from roughly 10 pixels full width at 10\% max to over 40 pixels.  This means the beam footprint as the light passed through both QWP and HWP retarders was substantially different.  Figure \ref{psfs_polarimetric_manystars} showed the spatial profiles for four stars to illustrate the change in telescope focus.  This defocus changes the size of the beam and also where the beam from each field angle passes through the optic.

There are two common methods to subtract this fringing.  First, an unpolarized standard star is observed under the same conditions and the corresponding spectral fringe is subtracted from the science target spectra.  Second, Fourier filters can be applied to the science target spectra without any need for a calibration target provided the frequencies to be filtered are known for the specific instrument configuration.

In the first method, the calibration standard can be observed with very high SNRs and simply subtracted without degrading the SNR of the target. This assumes that the spectral fringe is stable between calibration standard and science target.  One potential error is that telescope guiding drifts can change the angle and location of the beam through the optic. Another is that gravity or time dependent instrument changes (flexure, temperature, focus, etc) can substantially alter interference fringes. 

\begin{figure} [!h, !t, !b]
\begin{center}
\includegraphics[width=0.95\linewidth, angle=0]{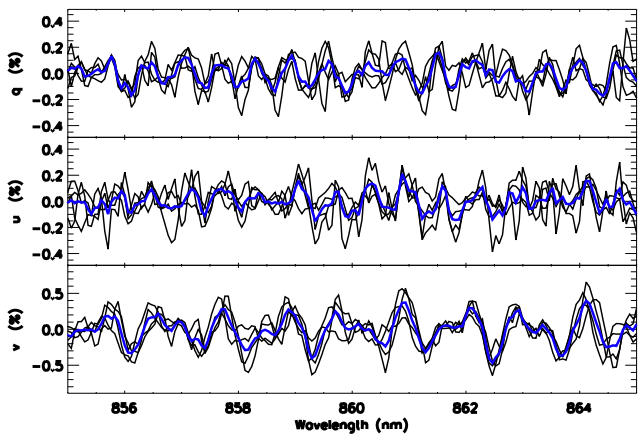}
\caption{ \label{ripple_manystars_CrH_short} The spectral fringes present in 4 observed calibrator targets for the 855nm to 865nm wavelength region. The unpolarized standard HD 174160 from August 23rd had a SNR of 1200,  the unpolarized but magnetic HD 20630 star from August 22nd had a SNR of 1000 and August 23rd had an SNR of 2000.  Ceres, though presenting continuum polarization, matches the unpolarized standards from August 23rd with an SNR of 2000. The blue curve shows the average of all 4 individual spectra.}
\end{center}
\end{figure}

Figure \ref{ripple_manystars_CrH_short} shows the $quv$ spectra for the four possible calibration standards we observed August 22nd and 23rd.  A clear and repeatable spectral fringe signature present.  Ceres had a of SNR of 2000.  The two unpolarized but magnetic HD 20630 observations had SNRs of 1000 on August 22nd and 2000 on August 23rd.  The unpolarized standard HD 174160 had a SNR of 1200. Though Ceres is an asteroid with a continuum polarization, it is not expected to show significant spectropolarimetric signatures in this wavelength region.  Since the telescope focused changed substantially between the unpolarized standards and Ceres, some systematic variation is expected between the observations. There is some small but statistically significant variation between the fringes detected in Figure \ref{ripple_manystars_CrH_short}. We note that the Ceres spectra are very similar to all the unpolarized standard observations.  Additionally, the Ceres spectra were only mildly defocused and match the brown dwarf focus position much closer than the unpolarized standards.

\begin{figure*} [!h, !t, !b]
\begin{center}
\includegraphics[width=0.45\linewidth, angle=0]{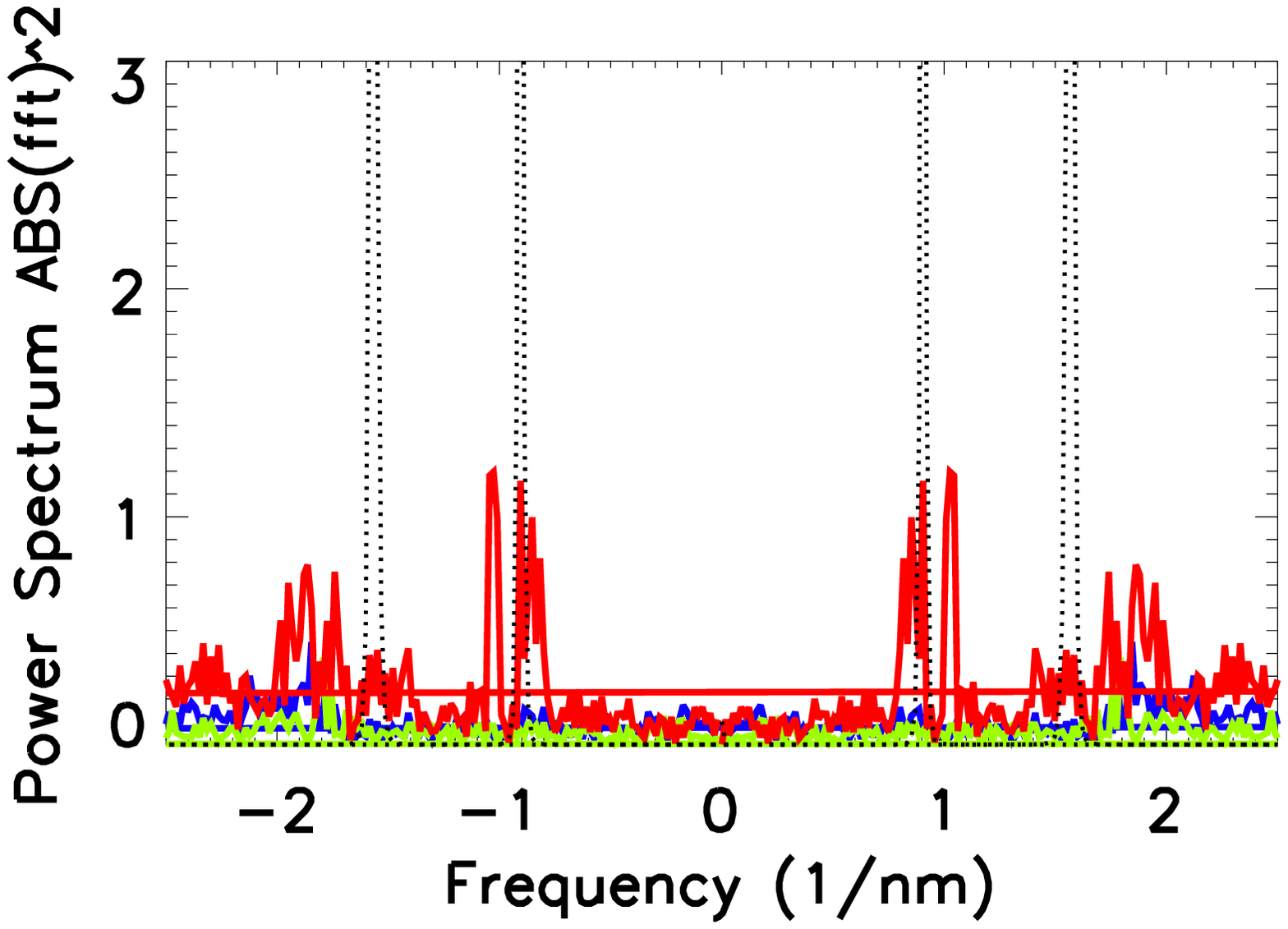}
\includegraphics[width=0.45\linewidth, angle=0]{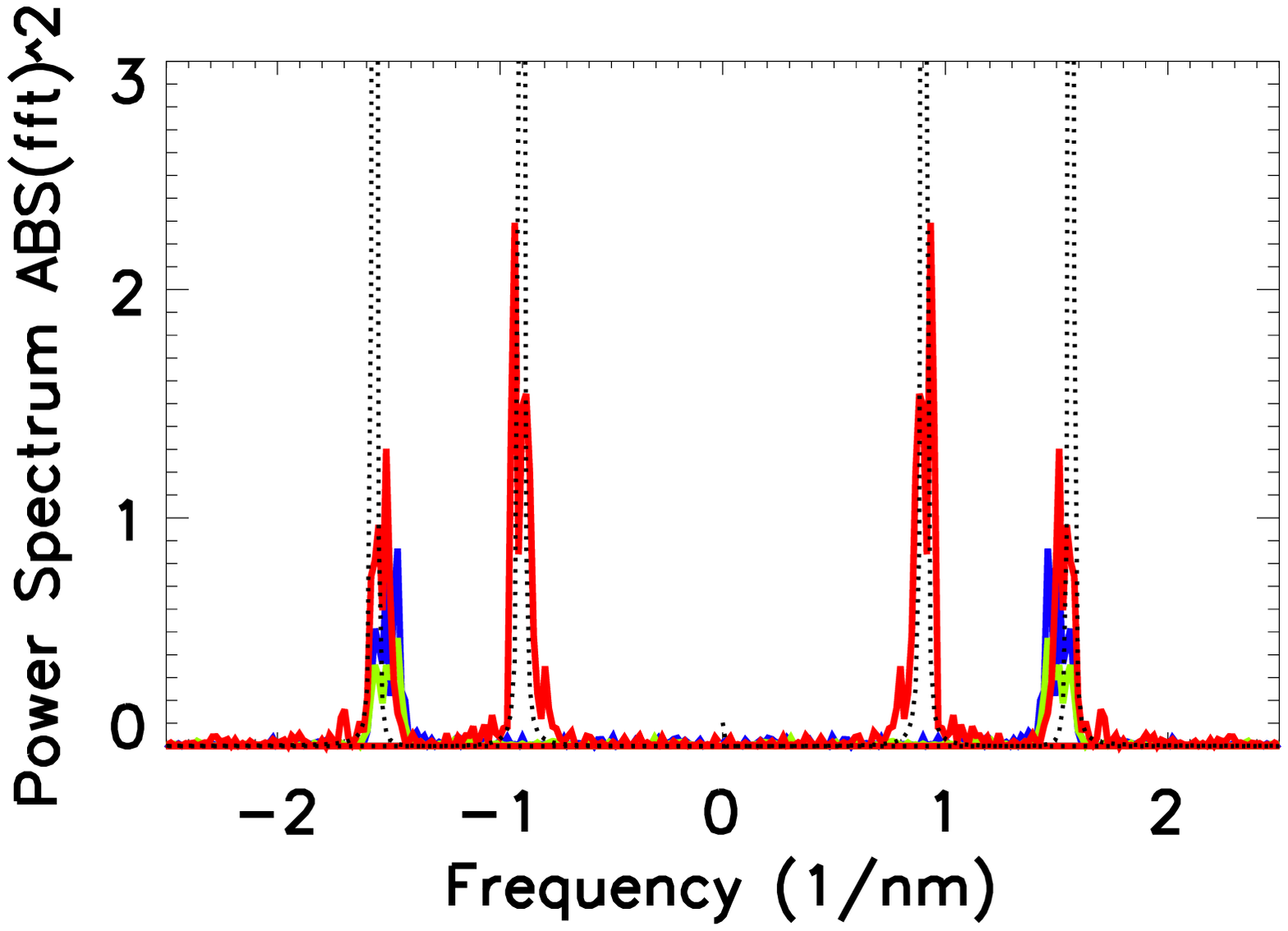}
\includegraphics[width=0.45\linewidth, angle=0]{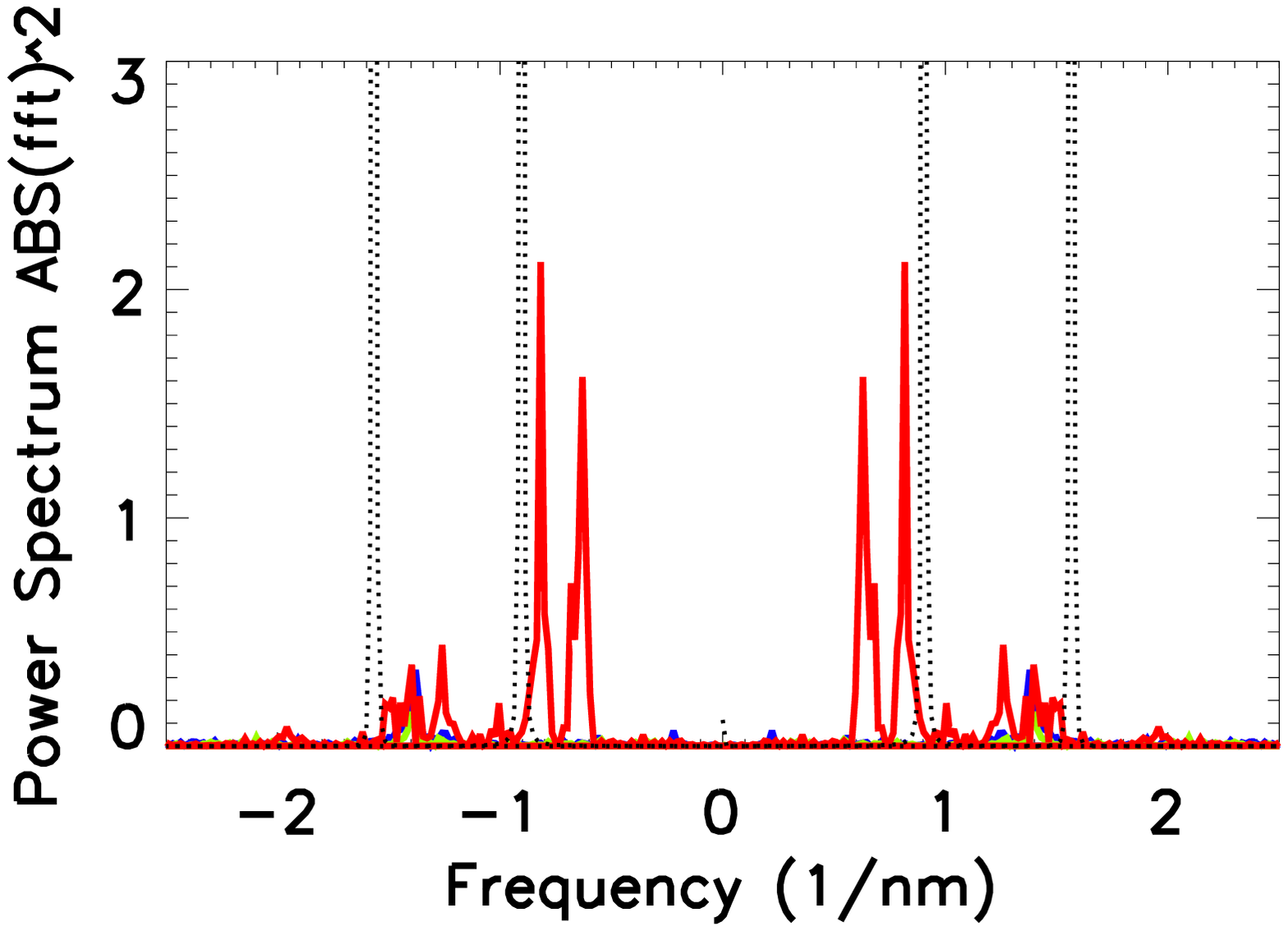}
\includegraphics[width=0.45\linewidth, angle=0]{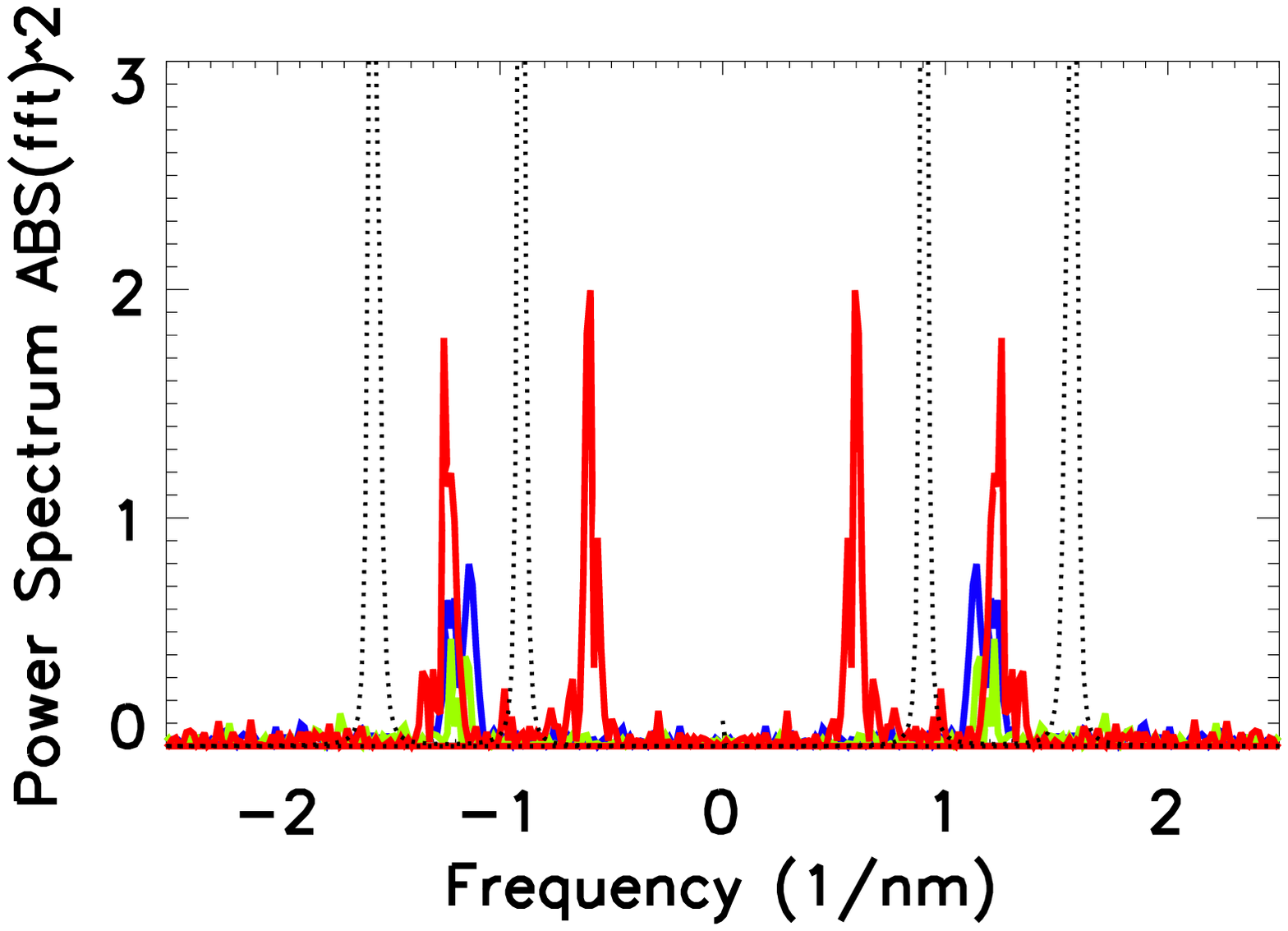}
\caption{ \label{ev_lac_fft_power_chromatic} The EV Lac power spectra for the $quv$ fringe in select spectral bandpasses. The central wavelengths are 817nm, 875nm, 934nm and 992nm with 1000 spectral pixels per computation. Blue shows Stokes $q$.  Green shows Stokes $u$.  Red shows Stokes $v$.  The Stokes $v$ power spectra in red show multiple peaks that have significant frequency width.  The dashed lines show a simple model with two cosine functions multiplied together to give a measure of the single-frequency peak width. The cosine functions have periods of 0.33 nm$^{-1}$ and 1.23 nm$^{-1}$ which arise to the sum-frequency of 1.56 nm$^{-1}$ and the difference-frequency of 0.90 nm$^{-1}$ in the power spectrum.  }
\end{center}
\end{figure*}

The second method for removing these spectral fringes involves computing the power spectrum of each individual Stokes parameter and then filtering unwanted frequencies. We compute the power spectrum conventionally as ABS(FFT($quv$))$^2$. 

The $quv$ fringe has significant wavelength dependence in the 800 to 1000nm region.   We choose 817nm, 875nm, 934nm and 992nm to illustrate the wavelength dependence.  We use small wavelength intervals of 1000 spectral pixels centered on the bandpasses to illustrate the changes with wavelength. Figure \ref{ev_lac_fft_power_chromatic} shows the power spectra for Stokes $q$, $u$, and $v$ in all four wavelength bandpasses. 

There are typically multiple frequency components, particularly in the Stokes $v$ measurements. For instance, the Stokes $qu$ spectra have substantial power at a 1.56 nm$^{-1}$ period in the 875nm bandpass.  The Stokes $v$ spectra are more complex in the same bandpass with power at both 1.56 nm$^{-1}$ and 0.90 nm$^{-1}$ periods. If you model this $v$ power spectrum as a multiplication of two cosine functions, you get periods of 0.33 nm$^{-1}$ and 1.23 nm$^{-1}$ which arise to the sum-frequency of 1.56 nm$^{-1}$ and the difference-frequency of 0.90 nm$^{-1}$. This simple model is shown as the solid black curves in every panel of Figure \ref{ev_lac_fft_power_chromatic}.  Since the Stokes $v$ data has two retarders mounted in the beam, complex interactions are expected. 

The wavelength dependence of the fringe is most readily seen as the change in period of the oscillations. The relative strength of the fringe at various periods also changes.  At 817nm, the fringe for Stokes $v$ is largely contained at periods of 1 nm$^{-1}$ and 2 nm$^{-1}$ but with two separate frequency peaks seen at each period.  By 875nm, the Stokes $v$ power spectra frequencies have shifted to shorter frequencies and there are two separate but identifiable peaks within each main frequency band.  At longer wavelengths of 934nm, there are two very clear independent frequencies smaller than 1 nm$^{-1}$ and two much smaller peaks around 1.5  nm$^{-1}$.  The fringe variation in $q$ and $u$ in the 934nm bandpass is barely detectable.   However, in the 992nm bandpass, the $q$ and $u$ variability are clearly detectable again. 

We compared the Fourier filter method with a direct subtraction of a stellar calibration observation and find good agreement where we have sufficient SNR in the Ceres spectrum. However, the method if directly subtracting standard star observations under the same optical configuration seems to be robust and to deliver higher SNRs when calibration standards are observed at very high SNRs.   

\begin{figure} [!h, !t, !b]
\begin{center}
\includegraphics[width=0.99\linewidth, angle=0]{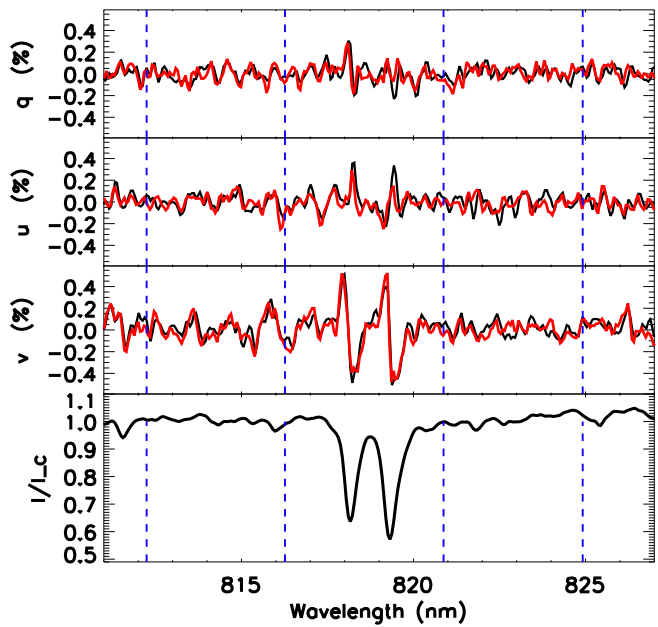}
\caption{ \label{evlac_spectral_ripple_comparison} A comparison of the EV Lac $quv$ spectra after fringe subtraction using different combinations of stellar and standard observations. The red curve shows the EV Lac spectra after subtraction of all flexure-corrected stellar observations (all stars and calibrators).  The black curves show the EV Lac spectra when using only bright star and calibrator observations (V 2054 Oph, Ceres and unpolarized stars HD 20630 and HD 174160, no brown dwarfs).  The dashed blue lines define two regions used for computing SNR statistics as well as continuum normalization for the intensity profile.  The blue continuum runs from 812.2nm to 816.3nm while the red continuum runs from 820.9nm to 824.9nm covering 70 spectral pixels in each band.}
\end{center}
\end{figure}

Figure \ref{evlac_spectral_ripple_comparison} shows the EV Lac $quv$ spectra after subtraction of the spectral fringe using two different sets of calibrators.  We tested the fringe subtraction methodology by using different groups of observations to create an average spectrum at much higher SNRs.  In one group, we simply averaged all observations from Table \ref{Table_LRISp_Observations} after applying the various flexure and wavelength drift corrections to create an average fringe spectrum.  For a second fringe spectrum, we averaged only the non-brown-dwarf targets at high SNR observations from Table \ref{Table_LRISp_Observations} which excludes the LSRJ and 2MASS stars. These two spectra allow us to verify consistent results in effectively subtracting this spectral fringe.   

Figure \ref{evlac_spectral_ripple_comparison} shows that the resulting fringe subtracted EV Lac $quv$ spectra are essentially identical.  The SNR is estimated using the standard deviation in each $quv$ spectrum on both the blue and red continuum. The blue continuum runs from 812.2nm to 816.3nm while the red continuum runs from 820.9nm to 824.9nm with each band covering 70 spectral pixels outlined by the dashed blue vertical lines in Figure \ref{evlac_spectral_ripple_comparison}.  This method gives SNRs from 950 up to 1640 for the red curve and 1310 up to 1800 for the black curve.  As an independent test of the fringe subtraction, we apply a high-pass filter to the data by subtracting a 9-pixel boxcar smoothed spectrum from the fringe subtracted data sets. This 9 pixel smoothing width represents the full width 10\% max of the delivered optical instrument profile measured from arc line exposures.  These residual variations after filtering are dominated by shot noise and any residual cosmic ray damage.  The SNRs of these high pass filtered data sets are in the range of 1700 to 2200 showing effective removal of the fringes down to typical statistical limits of 0.05\% in any individual $quv$ spectrum.  

\subsection{Slit stepping for High SNR}

\begin{figure} [!h, !t, !b]
\begin{center}
\includegraphics[width=0.75\linewidth, angle=90]{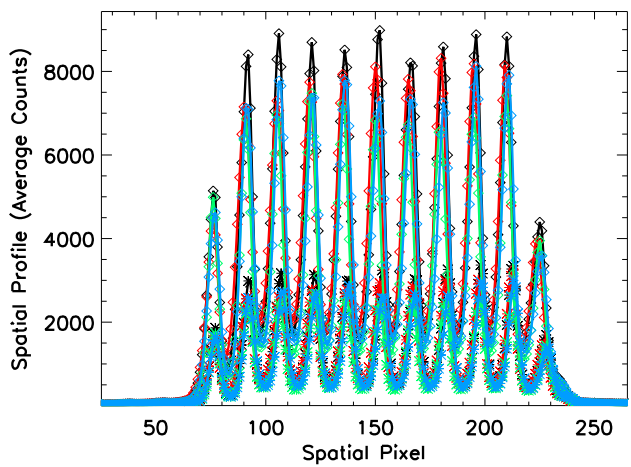}
\caption{ \label{slit_stepping_psfs} An example of the slit-stepping script for achieving high SNR measurements.  The star was moved in 1 arc second increments along the length of the slit 10 times for a total of 11 samples of the local seeing and telescope jitter. Each color represents a different exposure.  One of the two polarized beams had substantially higher throughput. Variations along the spatial pixels represents the changing flux on 1 second timescales. }
\end{center}
\end{figure}

	An additional method for increasing the SNR is to step the target along the slit length during an exposure.  An example of this procedure is shown in Figure \ref{slit_stepping_psfs}. The guider offset script was set to move the target along the slit in 1 arc second steps during an exposure.  A mode like this allows the user to achieve near-saturation brightness levels on the detector over an order of magnitude more spatial pixels without requiring readout of the detector.  For LRISp on bright targets, the CCD readout time is a major limitation. Readout can take up to a minute while the integration time to saturation is a small fraction of this time. By implementing this slit-stepping mode, the duty-cycle and efficiency of bright target observing campaigns can be kept high with integration times substantially longer than the readout time.

\begin{figure} [!h, !t, !b]
\begin{center}
\includegraphics[width=1.05\linewidth, angle=0]{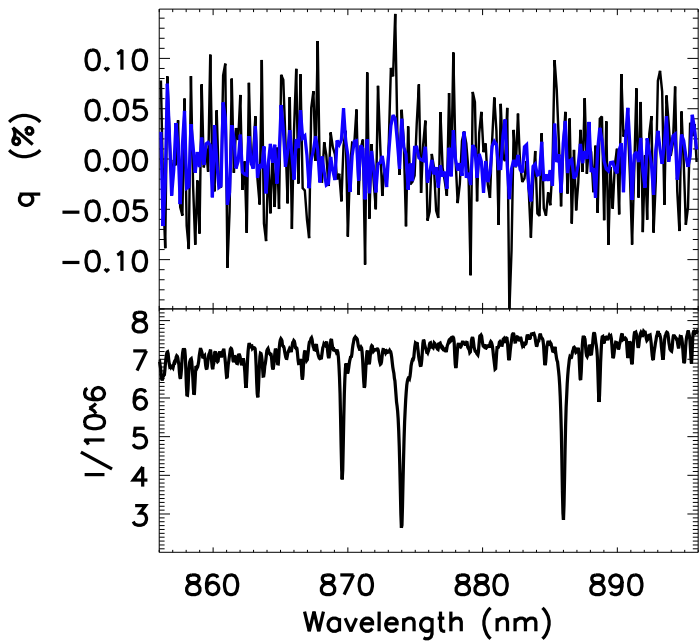}
\caption{ \label{slit_stepping_spectra} An example of the $q$ and intensity spectra for a field star using the new slit-stepping script.  This script allowed us to achieve high SNR measurements ($>$3000).  The star was moved in 1 arc second increments along the length of the slit following Figure \ref{slit_stepping_psfs} before readout, allowing for high duty-cycle observations of bright targets without saturation. The blue curve shows an average of 5 repeated exposures on this target with SNRs ranging from 1200 to 1500 each.}
\end{center}
\end{figure}

	This slit-stepping mode resulted in substantial change in the telescope beam footprint as the light passed through the retarders.  This changes the spectral fringe.  A simple Fourier filter as shown above is sufficient to remove spectral fringes.  An example of a high SNR Stokes $q$ and intensity spectrum for a bright field star HD345495 is shown in Figure \ref{slit_stepping_spectra}.  After fringe removal, the spectra were averaged spectrally by 4 pixels to show SNRs above 3000 at 1-point per resolution element spectral sampling. 

	We acquired 5 separate spectra of this field star using this slit-stepping mode.  The guider performance did result in some small variation in between steps, but the SNR's for each individual polarization measurement were between 1200 and 1500 at full spectral sampling.  After combining and spectrally averaging by a factor of 4, we achieve a final SNR of 4500 for a polarimetric sensitivity of 0.022\% as seen in the blue curve of Figure \ref{slit_stepping_spectra}. The higher SNRs achieved by spectral and temporal averaging demonstrates that the SNRs are statistically limited and not dominated by some systematic errors.

\subsection{Internal Calibrations}

The coordinate reference frame for linear polarization can be verified with optics mounted in the internal filter wheel \citep{Goodrich:2003kv}. The wheel includes two calibration polarizers in addition to the QWP used for circular polarization. The infrared polarizer is said to have a wavelength range of 750nm to 1050nm \citep{Goodrich:2003kv}.  We used this calibration optic to perform an independent assessment of the cross-talk between linear polarization states introduced by having chromatic fast axis orientation variations in the HWP.  

We follow a standard method to identify the HWP fast axis orientation as well as the overall degree of polarization delivered by the calibration optics. The flat field lamps are used to illuminated the fixed calibration IR polarizer.  We then take a series of exposures while rotating the HWP.  In our case, we rotated the HWP from -7$^\circ$ to 89$^\circ$ in steps of 2$^\circ$.  This allows us to have high angular sampling as well as recording both the minimum and maximum intensity transmitted through both beams of the polarizing beam splitter.  With 49 independent measurements at a range of HWP orientations, we can model the transmitted intensity as a simple function and fit for the calibration parameters. 

\begin{figure} [!h, !t, !b]
\begin{center}
\includegraphics[width=0.95\linewidth, angle=0]{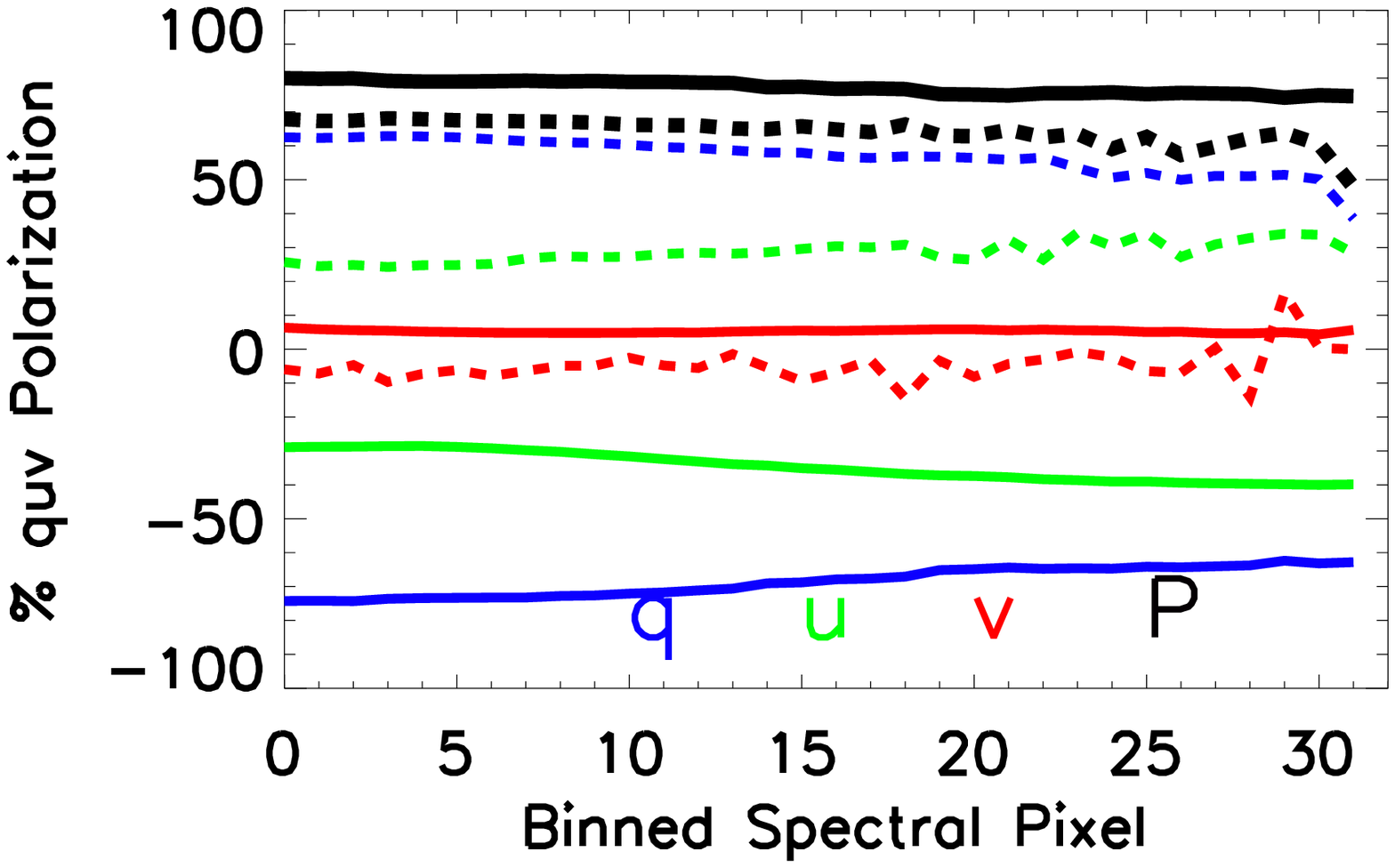}
\caption{ \label{polcal_polarizer_hwp_test} The degree of polarization for the IR calibration polarizer and the half-wave plate fast axis orientation. The measured degree of polarization (DoP) for the polarizer is shown on the right and y-axis. The DoP is above 95\% for wavelengths shorter than 900nm. At longer wavelengths the DoP falls to below 20\% at 1000nm. The fast axis orientation for the HWP is shown as the colored curves with the left-hand y axis. The colors correspond to the two orthogonally polarized beams recorded on the detector produced by the polarizing beam splitter. Each colored fast axis location has two separate curves corresponding to fitting a function with either a fixed angular modulation period ($P=90^\circ$) or with a variable period ($P$). The fast axis orientation changes by about 2.5$^\circ$ from 780nm to 950nm. Beyond 950nm, the calibration polarizer has low DoP and the fast axis orientations measured in the two beams starts to diverge. Additionally, the computed modulation period diverges from the nominal angle of $P=90^\circ$.  Likely the low DoP and other polarization artifacts from the polarizer itself cause the measurement technique to give errors. See text for details.}
\end{center}
\end{figure}

At every wavelength, we model the system as a combination of an imperfect polarizer and a HWP that rotates the polarization by some angle. We also include a fit for the period of the modulation function with HWP angle as a test on the accuracy of the model.  Imperfections in the assumptions about the polarizer behavior as well as errors in rotation stage encoders values can manifest as deviations. The functional form of the intensity modulation with HWP angle is represented as an unmodulated intensity ($I_0$) constant plus a modulated intensity ($I_m$) amplitude times the modulation function itself. For an ideal HWP with a varying fast axis orientation, the modulation is simply a Cosine function with an unknown orientation ($\theta$). The angular period of the polarization modulation ($P$) should be 90$^\circ$ for a perfect HWP.  The functional form is thus:

\begin{equation}
I = I_0 + I_m COS( 2\pi  \frac{\theta - \theta_0}{P} )
\end{equation}

As a test of the method, we perform fits both with a fixed $P=90^\circ$ and a variable period.  The results of the functional fit at every wavelength are shown in Figure \ref{polcal_polarizer_hwp_test}. We included fits where $P$ is both fixed and allowed to vary. The fast axis orientation variation is seen in color and varies by about 2.5$^\circ$ from 780nm to 950nm.  In the 950nm to 1050nm the fits begin to diverge.  The angular modulation period deviates from 90$^\circ$ and the two orthogonally polarized beams do not give the same location for the HWP fast axis. The assumption of the calibration polarizer being simply represented as a fractional polarization at a fixed angle and the behavior of the HWP appears to break down. 

However, this testing does show that the expected rotation of the linear polarization reference frame caused by deviations in HWP properties with wavelength is well controlled. The rotation of the plane of polarization for a science target should be easily compensated by use of a linear polarization standard star as is common for calibrations of LRISp \citep{Goodrich:2003kv,Goodrich:1995fg}.

\section{Daytime Sky Calibration Testing: Cross-talk}

As an independent test of the instrument calibration, we use the daytime sky polarization.  The quarter wave plate used to measure circular polarization is mounted in the calibration filter wheel.  As such, there are no calibration optics mounted in front of the instrument in the 2-retarder configuration. This presents a calibration challenge. Recently, we have been developing methods to use daytime sky polarization as a bright, highly polarized calibration source with a well known angle of polarization (AOP) to derive telescope properties \citep{Harrington:2011fz, Harrington:2010km}. We have developed algorithms to use this daytime sky polarization to compute the cross-talk introduced in the instrument while illuminating the entire optical train much more similar to star light. This method was successful in calibrating the AEOS telescope and the HiVIS spectropolarimeter where the linear to circular cross talk was 100\% at some wavelengths and telescope pointings.  This method provides an independent check of the internal calibration optics and can be used without relying on standard stars.  

\begin{figure} [!h, !t, !b]
\begin{center}
\includegraphics[width=0.95\linewidth, angle=0]{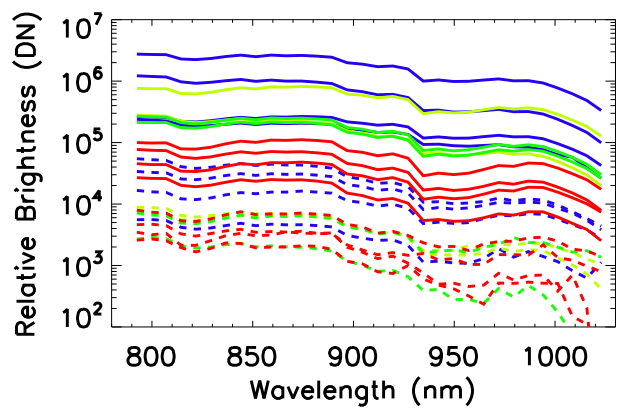}
\caption{ \label{sky_flux_vs_time} Daytime and twilight sky intensity measured with LRISp as the sun set on August 22nd.  The first data set was recorded with the telescope at an azimuth of 270 shown in solid lines.  The second data set was recorded at an azimuth of 180 shown in dashed lines. Each color shows a different modulation state. Note there is at least an order of magnitude change in brightness between spectra just due to the high degree of polarization in the daytime sky at the Zenith.  The brightness dropped by 3 orders of magnitude between the start and end of the test as the sun set. }
\end{center}
\end{figure}

During our observing run, we obtained permission to open the Keck dome before sunset. We observed the Zenith with LRISp before and during twilight on both August 22nd and 23rd.  During this time, the sky degree of polarization changes quickly and takes on spectral features in atmospheric absorption bands.  The shadow from the earth propagates upward through the atmospheric column during twilight. On August 22nd we observed the Zenith sky polarization with the telescope pointed at an azimuth of 270$^\circ$ and then 180$^\circ$.  On August 23rd, we observed the Zenith at azimuths of 360$^\circ$ and then 090$^\circ$.  Figure \ref{sky_flux_vs_time} shows the detected brightness for two complete polarimetric data sets on August 22nd.  

\begin{figure} [!h, !t, !b]
\begin{center}
\includegraphics[width=0.95\linewidth, angle=0]{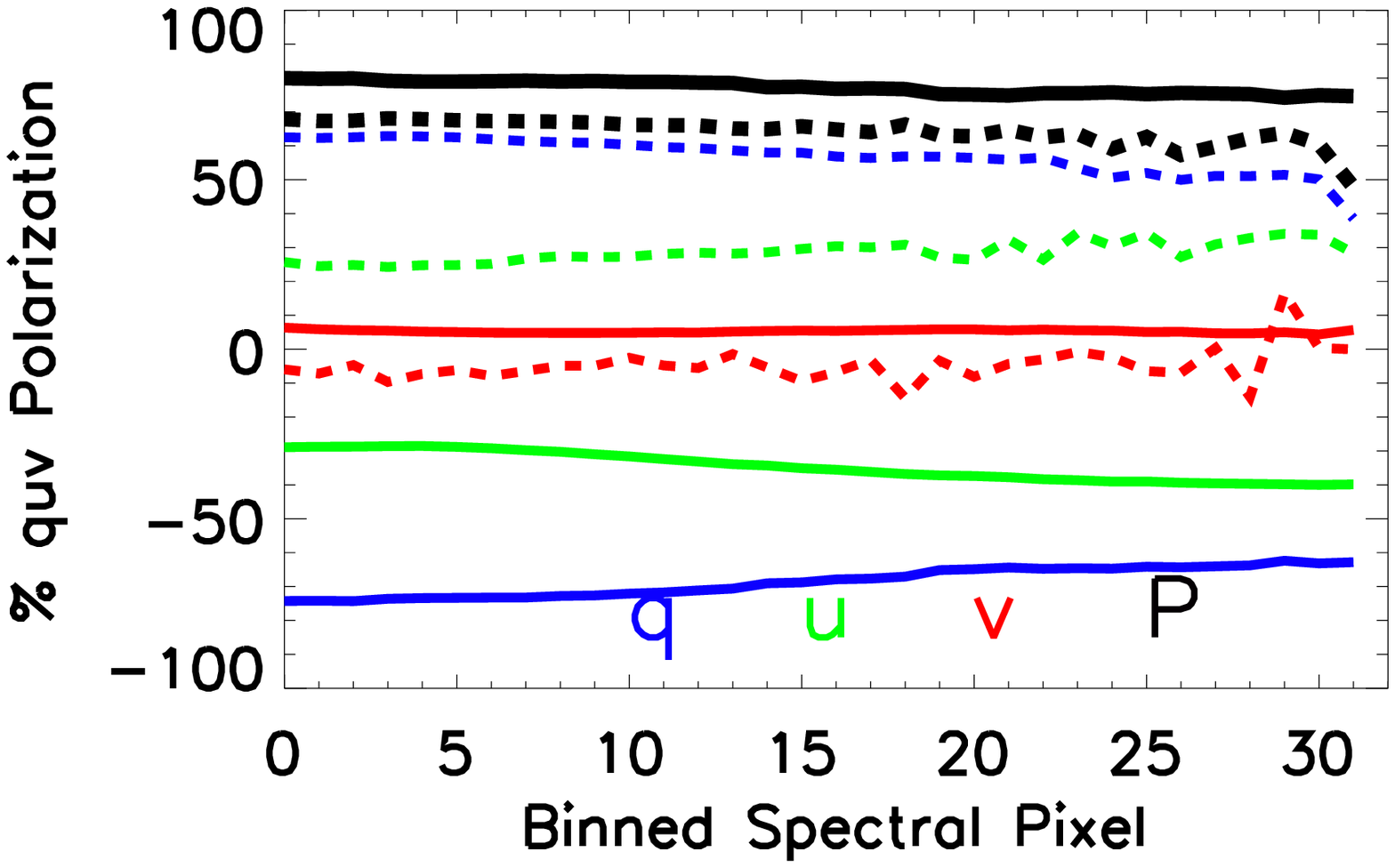}
\caption{ \label{sky_swp} The Stokes $quv$ spectra and the corresponding degree of polarization (DoP) measured with LRISp on August 22nd.  The solid lines show the first data set with the sun above the horizon and the telescope at an azimuth of 270. The dashed lines show the second data set with the telescope at an azimuth of 180 as the sun was setting. The Stokes $v$ measurement (dashed red line) has the lowest flux levels and the corresponding highest noise at longer wavelengths. Clear linear-to-circular cross talk is seen as the nonzero Stokes $v$ component of roughly 5\% changing sign as the instrument rotates.}
\end{center}
\end{figure}

At the Zenith on a mountain site with the sun near the horizon, we expect to measure degree of polarizations above 60\% as measured by all-sky imaging polarimeters common in the atmospheric sciences \citep{Swindle:2014ue,Swindle:2014wc, Dahlberg:2009jh, Dahlberg:2011wk, 2010SPIE.7672E..0AS, Pust:2006gc, Pust:2007fl, Pust:2009fq}.  Figure \ref{sky_swp} shows the Stokes $quv$ spectra and the associated Degree of Polarization (DoP) measured with LRISp on August 22nd. 

We measured 65\% to 70\% DoP in the first data set and 80\% in the second set.  This slight increase in DoP as the sun reaches the horizon is expected from standard Rayeligh scattering theory on a clear day. 

We can see that the Stokes $v$ measurements are small but nonzero.  This is expected as there is a known misalignment between the retarders and the polarizing beam splitter orientation in addition to the expected chromatic change in the QWP fast axis orientation \citep{Goodrich:1995fg}. Note that the QWP is fixed in the calibration wheel mount and optimization must be done manually in the present instrument configuration.  The daytime sky has no circular polarization measured to limits better than 1\% \citep{Swindle:2014ue,Swindle:2014wc}.  Our measurements show that the linear to circular cross-talk is roughly a few percent of the incoming linear polarization signal.  

We also note that the angular relationship was verified to be preserved following our algorithms outlined in \cite{Harrington:2011fz}.  On both August 22nd and 23rd we observed the Zenith with the telescope at cardinal pointings of North, East, South and West.  The angle between the measured Stokes vector and the theoretical Rayleigh sky vector was preserved to better than 5$^\circ$ as the instrument rotated on both days.  

The high measured degree of polarization for the daytime sky shown above at long wavelengths also suggests efficient modulation and polarizing beamsplitter functionality even though the internal calibration polarizer was unable to accurately verify long wavelength performance from the degraded internal polarizer performance.

\subsection{Intensity to Polarization Cross-talk}

Another critical test for any spectropolarimeter is to measure the level of cross-talk between the intensity spectrum and the polarization spectra. There are two separate types of intensity to polarization artifacts.  The instrumental induced polarization is the first type.  This real polarization signal is generally caused by the optics.  This is often called the {\it zero point} calibration and is typically measured with an unpolarized standard stars. This polarization should not reflect the incoming stellar line spectrum and is usually a smooth continuum function.  As with most spectrographs our unpolarized standard measurements show $\sim$1\% instrumental polarization as relatively smooth functions of wavelength.   

A problem with defocusing the telescope is that the defocused beam as masked by the slit creates unstable zero point instrument polarization when combined with the poor guiding performance of the Keck slit guider.  Defocusing as a means for increasing the dynamic range has several advantages for line polarization studies where the continuum is simply fit and subtracted.  However, the relatively unstable masking of the incoming beam breaks the circular symmetry of optical path and reduces the ability for the instrument to measure continuum polarization.  Care must be taken when defocusing the telescope as the continuum stability is reduced. In addition, the slit guider performance degrades at high airmass and continuum polarization is less stable when tracking targets at high elevations. 

A second kind of artifact is caused by instrument and data reduction artifacts.  This kind of intensity to polarization cross talk involves having the incoming stellar spectrum spuriously present in the polarization spectrum. Imperfections in detector linearity, background subtraction and other effects begin to cause limitations at high sensitivity levels c.f. \cite{1996SoPh..164..243K}.  In unpolarized standard stars, in addition to the instrumental continuum polarization, the stellar line spectrum along with it's first and second wavelength derivatives can be spuriously present. Careful correction for scattered light, ghost images, wavelength drifts, etc will only reduce this kind of $I$ to $QUV$ cross talk below some detection threshold. The dual-beam configuration in addition to beam-swapping during data reduction does minimize several artifacts to first order.  However, several second order effects are not completely removed and can easily dominate the $quv$ spectra at high SNR levels. 

\begin{figure} [!h, !t, !b]
\begin{center}
\includegraphics[width=0.95\linewidth, angle=0]{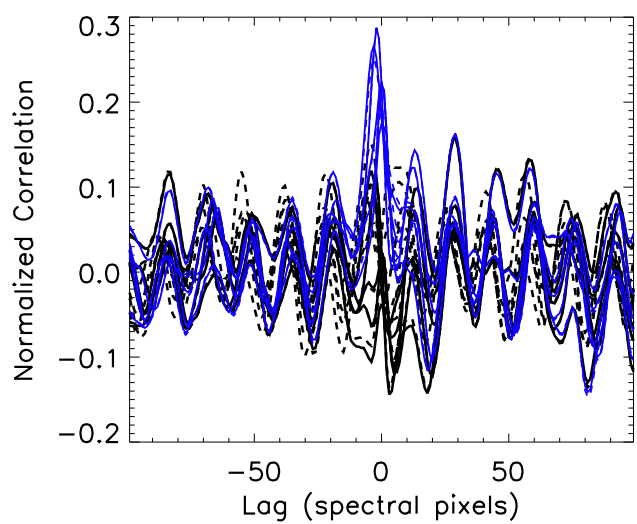}
\caption{ \label{i_to_quv} The cross correlation functions computed with the IDL C\_CORRELATE routine. The black lines show the correlation functions between intensity and the $qu$ spectra for the 5 slit-stepped spectra of HD 354495 described above. No substantial peaks are seen above the typical fluctuation levels of ~0.1.  The blue lines show the correlation functions when 0.02\% of the continuum normalized intensity spectrum was added to the individual polarization spectra. These blue correlation  curves show that a substantial correlation of 0.2 to 0.3 is present at zero lag giving rise to our upper limits for intensity to polarization cross-talk.}
\end{center}
\end{figure}

For LRISp, we took the five high SNR measurements in slit-stepped mode used to create Figure \ref{slit_stepping_spectra}. By running intensity-to-polarization correlations across many spectral lines, a clear signature of intensity to $quv$ spectra can be assessed.  Each of the five spectra had a SNR of 1200 to 1500 in each $q$ and $u$ Stokes parameter. A correlation analysis of 1000 spectral pixels between 856nm and 896nm yielded no significant I to $qu$ cross-talk at levels above the $qu$ SNR. The same test was run on the continuum normalized polarization profiles $Q$ = $I*q$ (sometimes differentiated as $q_c$ as opposed to $q$).  A test of the routines showed that 0.2\% of the intensity spectrum added in to either $qu$ or $QU$ spectrum was easily detectable with our correlation analysis. Figure \ref{i_to_quv} shows these normalized correlation curves computed with the IDL C\_CORRELATE function. A correlation of 1.0 would describe perfect correlation. The black curves show the correlation between the continuum normalized intensity spectra and the $qu$ spectra. The correlation curves fluctuate around zero with low amplitudes of 0.1 or less showing no substantial correlation.  The blue curves of Figure \ref{i_to_quv} show the correlation functions with the intensity added to the $qu$ spectra with a 0.2\% multiplier. Clear peaks are seen with correlations above 0.2 when we simulate intensity to polarization cross-talk at 2x the SNR levels (0.2\%).  This test sets our upper limits on the I to $qu$ cross-talk for our instrument configuration to $<$0.2\%.

\section{Comparison: LRISp Blue Channel}

	The blue channel of the polarimeter has some substantial differences in sensitivity to polarimetric errors when using common spectral extraction algorithms. We used the 680 dichroic and the 300/5000 grism during this campaign. The full-width half-max of arc line fits is between 6.0 and 7.5 spectral pixels showing substantial oversampling. There are slight differences of roughly 0.1 to 0.3 pixels FWHM between the two orthogonally polarized beams, showing a small but detectable difference in optical resolution between the orthogonally polarized beams.  The delivered spectral resolution is R$\sim$450 with 6 points per resolution element at 360nm. The resolution rises to R$\sim$790 at 7 point per FWHM sampling at 775nm.

\begin{figure} [!h, !t, !b]
\begin{center}
\includegraphics[width=0.95\linewidth, angle=0]{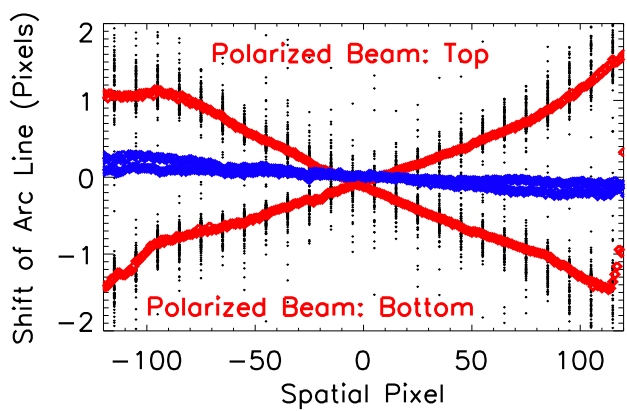}
\caption{ \label{spectral_tilt_comparison} A comparison between the pixel coordinate grid tilt geometrical calibrations derived using the arc lamp fits.  Black symbols show the computed wavelength center of several arc lines at selected spatial pixels on the CCD.  The red symbols show the median spatial location of all arc line wavelength centers for the two orthogonally polarized beams of the red channel.  Because the average wavelength solution in fixed ccd pixel coordinates was used, tilts between the orthogonally polarized beams are seen between a fixed pixel grid and the tilted arc lines.  This coordinate geometry must be corrected in the red channel by associating each spatial pixel with the correct wavelength derived from the arc lines as described above.  The blue channel has a greatly reduced tilt, as shown by the blue symbols for the median arc line location across the blue channel CCD. }
\end{center}
\end{figure}

	Unlike the red channel, the tilt of the monochromatic slit image is less than 0.1 pixels across 240 spatial pixels extracted. Tilted coordinate geometries can be overcome using up sampling and instrument calibrations, or even more complex instrument profile deconvolutions.  However, with such small spectral tilt in this blue channel, the wavelength errors and polarimetric artifacts are greatly reduced even when using simple extraction algorithms. The red and blue channel spectral tilts are shown in Figure \ref{spectral_tilt_comparison}. The red tilt is over 2 pixels across the slit image while the blue channel is more than an order of magnitude less.  Arc line exposures are used to determine how each wavelength falls across the CCD pixel grid.  The arc line central wavelength is mapped at every spatial and spectral pixel for every arc lamp line, shown as the black symbols in Figure \ref{spectral_tilt_comparison} only for the red channel.  The median spatial locations of all arc lamp lines along the CCD pixel grid are shown in red for the LRISp red channel.  The same median location for the blue channel is shown in blue. 

	The blue channel ccd is much thinner than the red channels deep depletion pixels.  The corresponding cosmic ray hit rate is greatly reduced.  The optimal extraction algorithms outlined for the red channel are always effective for rejecting cosmic ray hits as well as filtering some detector cosmetic issues. However, there is less need for this algorithm for modest exposure times.

\begin{figure} [!h, !t, !b]
\begin{center}
\includegraphics[width=1.05\linewidth, angle=0]{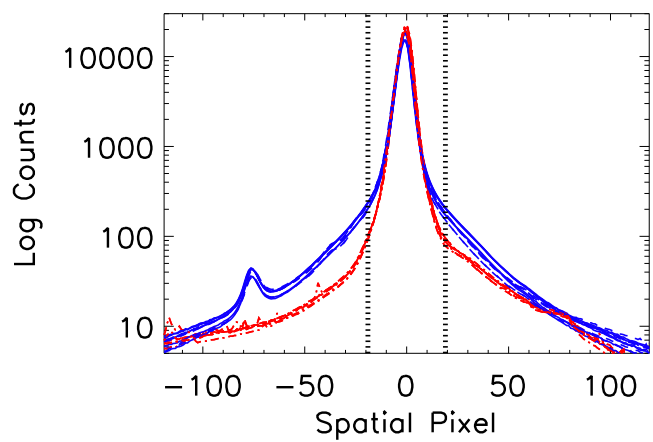}
\caption{ \label{evlac_psf_log} The spatial profiles computed for EV Lac.  The average spatial profile over all spectral pixels is shown for all 6 polarimetric exposures for the red channel (plotted in red) and the blue channel (plotted in blue). The blue channel has a ghost image and substantially wider scattered light in the wings of the spatial profile.}
\end{center}
\end{figure}

	Careful scattered light and background subtraction is more important in the blue channel.  There is a ghost image and substantially more scattered light width to the blue channel spatial profile. A comparison between the blue and red channels is shown in Figure \ref{evlac_psf_log}. The blue channel takes roughly double the number of spatial pixels to drop below the 1\% core flux level. Typical background subtraction algorithms will include some fraction of the stellar flux depending on the reduction choices made. This is easily subtracted along with night sky backgrounds provided this scattered light profile is known and properly accounted in the reduction parameter choices.  Night sky lines must be assumed to sit on top of a scattered light profile that extends substantially away from the core region, and scattered light backgrounds must be measured from an appropriate spatial distance away from the stellar core.
	
	The curve of growth is used to determine the optimal spatial extraction width when considering different noise source. We define the curve of growth as the computation of how much flux is included in the spectrum when setting progressively wider spatial profiles divided by the total detected flux. For instance, if the user picks a blue channel half width of 10 for a total of 21 spatial pixels in the blue extracted spectrum, the user is only including 90\% of the total flux. If the user would select a half width of 39 spatial pixels,  99.9\% of the light would be included.  In the red channel, care must be taken as cosmic ray damaged pixels are incompletely corrected and this noise source can dominate errors from incomplete flux inclusion.  Thankfully the scattered light is reduced in the red channel and correspondingly less spatial pixels are required to gather the majority of the detected stellar flux. 

\begin{figure} [!h, !t, !b]
\begin{center}
\includegraphics[width=0.99\linewidth, angle=0]{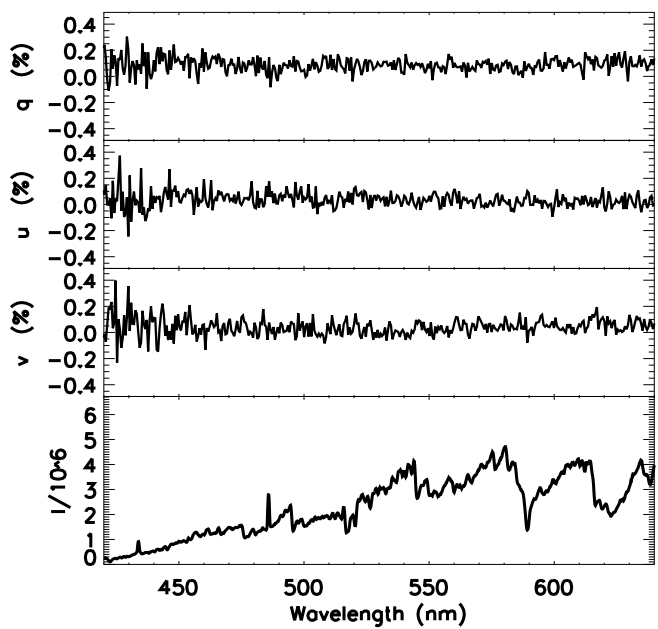}
\caption{ \label{evlac_signal_blue} The $quv$ and intensity spectra for EV Lac recorded on August 22nd 2012. The signal to noise ratios are estimated to peak above 2000 as seen in the $quv$ noise levels.}
\end{center}
\end{figure}

	Spectral fringe is not seen in the blue channel at these low spectral resolutions. An example high sensitivity polarimetric data set for EV Lac is shown in Figure \ref{evlac_signal_blue}. The signal to noise levels achieved are above 2000 per spectral pixel at full spectral sampling as estimated from the $quv$ spectra.  The $quv$ spectra are dominated by photon statistics across the entire 400nm to 700nm range.  The 680nm dichroic cutoff used did have some leakage beyond the cutoff wavelength, and spectra were extracted out to 775nm.  In the EV Lac spectra, a small spectral region was saturated and was subsequently not plotted.

\begin{figure} [!h, !t, !b]
\begin{center}
\includegraphics[width=0.99\linewidth, angle=0]{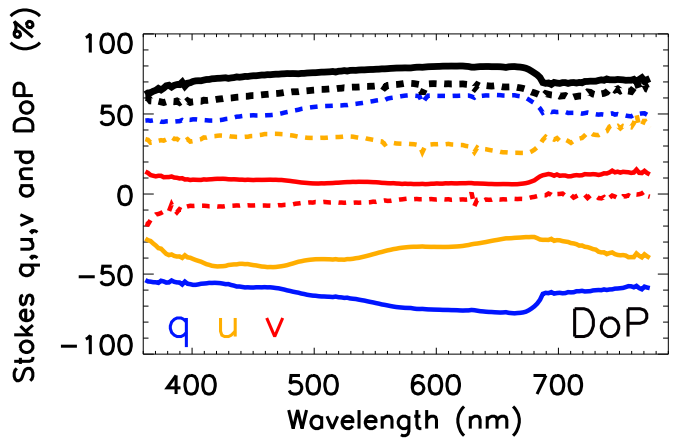}
\caption{ \label{evlac_sky_dop} The daytime sky polarization measured roughly 30 minutes before sundown on August 22nd 2012. Blue shows $q$, yellow shows $u$ and red shows $v$.  The total degree of polarization (DoP) is shown in black.  The 680nm dichroic cutoff greatly reduces the detected sky brightness for longer wavelengths.  Poor background subtraction and the blue channel scattered light profiles contribute to the change in polarization behavior given standard reduction algorithms.}
\end{center}
\end{figure}

	The blue channel showed similar cross-talk levels to the red channel given the standard retarder alignment and operating angles.  The daytime sky polarization observations were performed both on red and blue channels.  The measured sky degree of polarization is between 50\% and 80\% for the entire 360nm to 770nm range. Figure \ref{evlac_sky_dop} shows the daytime sky polarization measurements. Cross talk between linear and circular states is seen at similar levels to the red channel.

\section{Summary}

\begin{figure} [!h, !t, !b]
\begin{center}
\includegraphics[width=0.99\linewidth, angle=0]{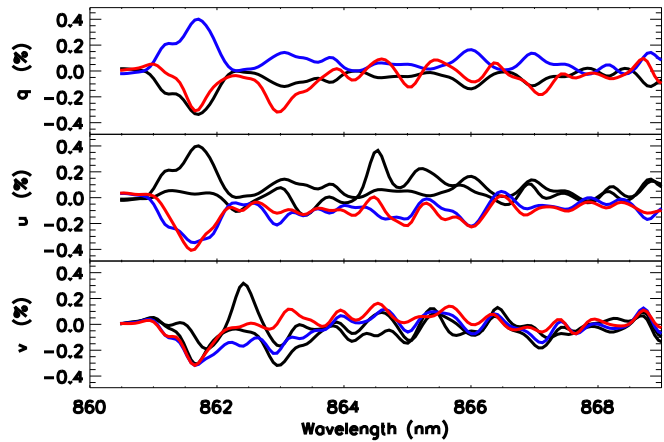}
\caption{ \label{magnetic_models} The model Stokes $quv$ signatures induced by a stellar magnetic field. Our brown dwarf target LSRJ was modeled under a range of magnetic field orientations, strengths and filling factors to show examples of the $quv$ morphology variation with magnetic field properties. Typical kilogauss fields as inferred from radio and optical measurements produce detectable signatures in the CrH band here with amplitudes of less than 0.4\%. }
\end{center}
\end{figure}

We have developed a data reduction pipeline that can calibrate the LRISp spectropolarimeter and achieve signal-to-noise ratios above 2000 limited by photon statistics at full resolution and sampling.  Spectral binning and temporal averaging achieved SNRs of 4500 without obvious visible instrumental artifacts.  We have demonstrated algorithms for overcoming several instrument artifacts and tested them on both red and blue channels. The major polarimetric limitations proved to be wavelength drifts from flexure, spectral fringes from the retarders and effective removal of cosmic ray contamination in the red channel deep depletion CCDs.  With these calibrations, we have successfully reproduced magnetic field measurements in atomic lines of M dwarfs such as EV Lac and found new magnetic signatures in molecular bands of M dwarfs and brown dwarfs.  Sensitive comparisons of low resolution spectropolarimetric data with new magnetic field models such as Figure \ref{magnetic_models} can now begin.  With magnetic fields producing signals at the $>$0.2\% levels, we must achieve shot-noise limited performance to sensitivity levels substantially below this level as demonstrated here. 

The instrument flexure in addition to beam wobble from rotating retarders introduce wavelength instabilities. The drifts are a substantial fraction of a pixel within and between modulated polarimetric exposures. These systematic errors produce spurious instrumental artifacts that are proportional to the first and second derivative of the spectral intensity with wavelength, even if the tilted coordinate geometry is calibrated using standard arc lamp exposures. These artifacts exactly resemble signatures from stellar magnetic fields and must be effectively suppressed to achieve accurate science results. Using correlation techniques with telluric lines and atmospheric sky-glow lines we can effectively track and remove these wavelength drifts. 

The spectral fringes introduced by the retarders have amplitudes of 0.2\% in Stokes $qu$ and over 0.5\% in Stokes $v$ in the red channel. These fringes display wavelength dependent behavior in both amplitude and frequency for individual Stokes parameters. Simple Fourier filters can remove the fringes provided a bandpass of 50nm or less is used.  The Fourier filtering method gives very similar results to subtracting a calibration standard star observation.  However, calibration of fringes by subtraction of standard star observations does give better results.

Cosmic ray hits are present in more than half the spectral pixels for each Stokes parameter in a typical long exposures of the red channel on a faint target. Optimal spectral extraction techniques are an effective filter of this noise source. An estimate of the local spatial profile computed using adjacent wavelengths gives information about the expected profile at any individual wavelength. This local spatial profile is used in an iterative loop to identify and reject cosmic ray hits and to correct the damaged pixels using the best-estimate of the spatial profile shifted and scaled to each individual extracted wavelength.

We have applied our daytime sky based polarization calibration techniques to LRISp to derive estimates of the instrument $quv$ cross-talk and polarization reference frame. There is a few percent linear to circular cross talk present as seen by observed Stokes $v$ when looking at a $\sim$100\% linearly polarized daytime sky in both blue and red channels. This is expected given the wavelength dependent properties of the retarders and small misalignments in the retarder rotation stages.  We also assessed the intensity to $qu$ cross-talk on high SNR observations and derive upper limits.  The red channel $I$ to $qu$ and $QU$ cross-talk is below 0.2\% and limited by shot noise. 

This pipeline will allow high SNR use of LRISp and Keck on faint targets now that the major sources of instrumental error have been identified and suppressed.

\section{Acknowledgements}
We thank the Keck staff, support astronomers and in particular Dr. Bob Goodrich and Dr. Hien Tran for their support in operating the telescope during daylight hours and in developing scripts for new slit stepping modes. Dr. Harrington and Dr. Berdyugina acknowledge support from the InnoPol grant: SAW-2011-KIS-7 from Leibniz Association, Germany, and by the European Research Council Advanced Grant HotMol (ERC-2011-AdG 291659).  Dr Berdyugina acknowledges the support from the NASA Astrobiology Institute and the Institute for Astronomy, University of Hawaii, for the hospitality and allocation of observing time at the Keck telescope.  Dr. Kuhn acknowledges the NSF-AST DKIST/CryoNIRSP program. This program was partially supported by the Air Force Research Labs (AFRL) through salary support for Dr. Harrington. This work made use of the Dave Fanning and Markwardt IDL libraries.

\bibliographystyle{aa}
\bibliography{ms_pasp_submitted_2}

\end{document}